\documentclass[review]{elsarticle}

\usepackage{lineno,hyperref}
\usepackage{xcolor}
\usepackage{natbib}

\newcommand{\degree}{\ensuremath{^\circ}}

\journal{Progress in Surface Science}

\begin{document}

%%\linenumbers

\begin{frontmatter}

\title{Doped and Codoped Silicon Nanocrystals:  the Role of Surfaces and Interfaces}
\author{Ivan Marri, Elena Degoli}
\author{Stefano Ossicini\corref{mycorrespondingauthor}}
\cortext[mycorrespondingauthor]{Corresponding author}
\address{Dipartimento di Scienze e Metodi dell'Ingegneria, Universit\'a  di Modena e Reggio Emilia,Via Amendola 2 Pad. Morselli, I-42122 Reggio Emilia, Italy and CNR-Istituto di Nanoscienze-S3, via Campi 213 A,I-41125 Modena, Italy}
\ead{stefano.ossicini@unimore.it}

\begin{abstract}
Si  nanocrystals  have been extensively studied because of their novel properties and their potential applications in electronic, optoelectronic, photovoltaic, thermoelectric and  biological devices. These  new  properties are achieved through the combination of the quantum confinement of carriers and the strong influence of surface chemistry.  As in the case of bulk Si the tuning of the electronic, optical and transport properties is related to the possibility of doping, in a controlled way, the nanocrystals. This is a big challenge since several studies have revealed that doping in  Si nanocrystals differs from the one of the bulk. Theory and experiments have underlined that doping and codoping are influenced by a large number of parameters such as size, shape, passivation and chemical environment of the silicon nanocrystals. However, the connection between these parameters and dopant localization as well as the occurrence of self-purification effects are still not clear. In this review we summarize the latest progress in this fascinating research field considering free-standing and matrix-embedded Si nanocrystals both from the theoretical and experimental point of view, with special attention given to the results obtained by ab-initio calculations and to size-, surface- and interface-induced effects.

\end{abstract}


\end{frontmatter}

\tableofcontents

\section{Introduction}
\label{sec:intro}

Silicon has become the most studied material in the past decades owing to its unique characteristics: Si is the second most abundant element (after oxygen) in the Earth’s crust, making up 25.7\% of its mass; it can be produced with impurity levels of less than
$10^{-9}$; it remains a semiconductor at higher temperatures than germanium; its native oxide is easily grown in a furnace and forms a better semiconductor/insulator interface than any other material. Silicon is not only easy to handle and fairly simple to manufacture but it is also cheap. Moreover it shows excellent electrical properties, high stability, optimal thermal characteristics and high mechanical strength. For these motives, this environmental friendly material is the most used element  in semiconductor industry and, today, it represents the electronic material per excellence. It is also commonly used for solar or photovoltaic applications and its role in the optoelectronic industry is becoming more and more important. 

However, the rapid evolution in the electronic, optoelectronic and photovoltaic sectors is facing severe constrictions due to the actual physical limits of Si-based technologies\cite{KimerlingASS2000,Ossicinibook2003,PrioloNat2014}. Some examples are: i) the limitations in the operating speed of microelectronic devices due to interconnects, ii) bulk Si is an indirect band-gap material which emits in the infrared region with very low efficiency, iii) radiative recombination times in Si are very large, so the de-excitation dynamics are strongly affected by the occurrence of fast non-radiative recombinations mechanisms, iv) bulk Si shows a significant free carrier absorption and Auger recombination rates which impede population inversion and, hence, optical gain, v) its band gap impedes fully exploitation of solar radiation in photovoltaic applications, vi) the absence of a bulk second-order dipolar nonlinear optical susceptibility due to the bulk crystal centrosymmetry, that do not permit to use bulk Si in order to produce a wide variety of wavelengths from an optical pump.

The increasing demand for new, innovative and more efficient devices has driven scientists to explore new functionalities in Si-based materials. In photonics the main interest lies in the possibility to merge electronics and photonics on the same chip and therefore to render Si a good light emitter. Moreover the introduction of second-order nonlinearity by proper material engineering would be highly desirable, because the all-optical data management requires nonlinear silicon photonics \citep{CazzanelliNatMat2012}. Finally, for photovoltaic applications we have to add to Si new features in order to maximize solar radiation harvesting and to minimize loss by thermalization processes. 

The advent of nanoscience and nanotechnologies and the consequent scaling down of Si structures to nanometer sizes have opened new possibilities to overcome the inability of bulk-Si as efficient light emitter. Moreover, the capacity of altering energy gap and optoelectronic properties by size reduction, and therefore to enhance solar light harvesting, has opened new possibilities in the development of new efficient, nanostructured, Si-based solar cell devices. 

\begin{figure}
\begin{center}
\includegraphics[width=11.33cm,height=5.67cm]{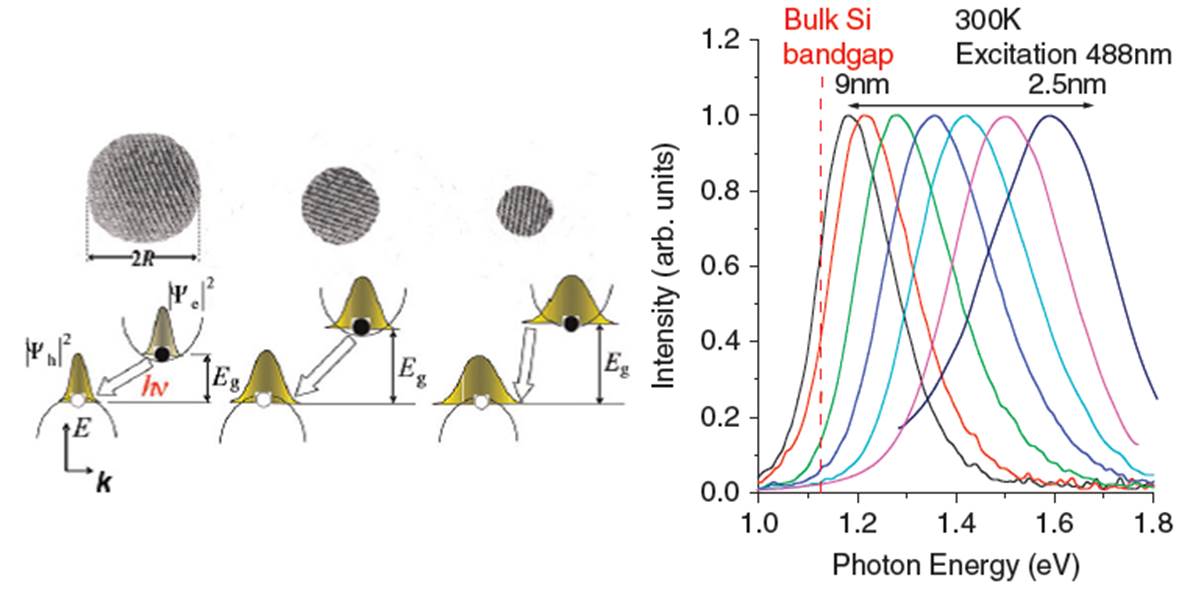}
\caption{(Color online) Right: quantum confinement effects in Si-NCs: scheme of electron and hole wave function spreads in the $k$ space. Reducing the radius R of the Si-NC leads to an opening of the band gap and to an increasing of the weight of quasi-direct transitions. Reprinted with permission from Ref. \citep{PelantPSSA2011}. Left: PL spectra of Si-NCs embedded in SiO$_2$  thin films at room temperature. The average diameters are changed from about 9 to 2.5 nm. Reprinted with permission from Ref. \citep{Fujiibook2010}}.
\label{fig:figure1}
\end{center}
\end{figure}

It has been suggested that the problems related to the indirect band-gap of bulk-Si might be overcome in highly confined systems.  For example, in low-dimensional Si-based nanostructures such as porous-Si, a quantum sponge made up of Si nanostructures, nanowires (Si-NWs) and nanocrystals (Si-NCs), the possibility to achieve efficient visible photoluminescence (PL) has been demonstrated  \citep{CanhamAPL1990,TakagiAPL1990,OssiciniPRL1994,BisiSSR2000}.  In these nanostructures the low-dimensionality causes the zone folding of the conduction band minimum of bulk Si, that is located near the $X$-point, thus introducing a quasi-direct band gap. Moreover the quantum confinement (QC) effect associated with the reduced size of the nanosystems enlarges the energy band gap enabling light emission in the visible range \citep{PelantPSSA2011}. This effect can also enhance the spatial localization of electron (e) and hole (h) wave functions and their overlap, and therefore the probability of e-h recombination (see Fig.~\ref{fig:figure1}).

Among the different Si-based nanostructures, Si-NCs have attracted, from the beginning, much interest because they exhibit very bright visible PL (see Fig.~\ref{fig:figure1}), which is tunable with respect to the dimension of the Si-NCs \citep{Fujiibook2010}, and sample dependent high quantum yields, ranging from 1$\%$ up to 60$\%$. The quantum yields can be enhanced by careful surface passivation and by avoiding oxygen contamination during all stages of the Si-NC preparation \citep{JurbergsAPL2006}. Moreover optical gain in Si-NCs has been successfully observed and discussed \citep{PavesiNature2000,KhriachtchevAPL2001,NayfehAPL2002,NegroAPL2003,RuanAPL2003,LuterovaAPL2004,
DohnalovaNJP2008,DegoliCR2009} as well as giant Raman gain \citep{SirletoNatCom2012,NikitinAPL2013}. Finally carrier multiplication effects in excited carrier dynamics after the absorption of a single high-energy photon have been observed in ensemble of Si-NCs, proving thus the possibility of better exploiting processes of photo-generated carriers to increase solar cell performances \citep{BeardNL2007,TimmermannNatPhot2008,TrinhNatPhot2012,GovoniPRB2011,GovoniNat2012,
MarriJACS2014,MarriBJN2015,MarriSE2016}. 

The possibility to exploit the size, shape and surface termination \citep{GregorJPCM2014} of the Si-NCs togheter with their low toxicity and good biocompatibility has prompted their application in several broad fields such as microeletronics
 \citep{Ossicinibook2003,TalapinCR2010,GuptaNANO2011}, photonics \citep{DaldossoLPR2009,KriaIJP2012}, non-linear optics \citep{ImakitaOE2009,BisadiPSSA2015,MakarovNL2017}, secure communications \citep{BisadiPSSA2015},
photovoltaics, in the so-called third-generation solar cells, solar fuels   \citep{PrioloNat2014,GreenPPRA2001,NozikNL2010,ConibeerPPRA2011,MangoliniJVST2013,LoperPSSA2013,SummonteSESC2014}, thermoelectrics \citep{DresselADMA2007,ClaudioPCCP2014} and biomedicine \citep{FarrJNanom2006,ErogboboACSNANO2008,ChengCSR2014,OstrovskaRSC2016,McveyCPC2017}.

Visible PL in Si nanostructures has been attributed to transition between states localized inside the Si-NC  \citep{KanemitsuTSF1996,DorigoniPRB1996,MoskalenkoPRB2007,SykoraPRL2008}, or between defects and/or interface states \citep{AllaPRL1996,WolkinPRL1999,IwayamaSSE2001,MartinNL2008,DovratPRB2009,KoponenPRB2009}.
Despite the discussion on which of the above mechanisms primarily determines the emission energy is still open, recent results seem to indicate the coexistence of both the effects, whose relevance depends on the structural properties of the sample \citep{AllaPRL1996,WolkinPRL1999,FilonovPRB2002,LuppiJAP2003,DaldossoPRB2003,ZhouNL2003,OssiciniJNN2008,HaoNanoT2009,
GourbilleauJAP2009,GuerraPRB2009,GuerraDegPRB2009,PulciSM2010,GuerraPRB2010}. In particular, it was suggested that for diameters above a certain threshold (from about 3 nm up to diameters of the order of two times the bulk Si exciton Bohr radius, about 9 nm) the emission peak of the Si-NCs simply
follows the QC model, while interface states assume a crucial
role only for small-sized Si-NCs (less than 3 nm) \citep{GuerraDegPRB2009,GuerraPSSB2010,GuerraPRB2011,SeguiniAPL2013}.  

It is well known that an additional degree of freedom in semiconductor materials design is given by the introduction of impurities. Controlled doping is at the heart of the modern Si-based semiconductor industry. The best known example is the p-n junction. Here elements of the column III and V, like the B and P atoms, are used to electrically dope the Si by introducing positive holes or negative electrons. 
The presence of the doping atoms changes remarkably the optical and electronic properties of the hosting material. B and P impurities introduce very shallow levels in the band gap of bulk Si, above the valence band and below the conduction band, than can be efficently ionized at room temperature increasing dramatically the conductivity of bulk Si. 

Doping in Si-NCs, as in bulk semiconductor devices, can be used to alter in a controllable way the electronic, optical and transport properties of nanomaterials. As a consequence, intentional doping with n- and p-type impurities can be exploited to design and realize novel devices based on Si-NCs. 

The possibility of enhancing the electrical conductivity of nanosized systems has been attempted, for instance, by obtaining porous Si from n- or p-doped bulk Si by means of an electrochemical etching \citep{CullisJAP1997}. Nevertheless, even for the larger mesoporous samples, a very low conductivity
was measured, despite the fact that the etching process
does not remove the impurities from the system \citep{PolisskiPB1999}. This
suggests that the ionization of the impurities at room temperature
may be strongly quenched with respect to the bulk. Therefore
the possibility of generating free charge carriers from
impurity states can be limited by size effects. 

Regarding the optical properties, the introduction in the Si-NC of an isoelectronic impurity can circumvent the indirect gap behavior of bulk Si. However Si does not possess proper isoelectronic impurities that can strongly localize excitons at room temperature and enhance the PL intensity. An alternative approach is given by codoping. i.e. the formation of Si nanocrystals with the same number of n- and p-type impurities \citep{FujiiAPL2005}. The first attempts to dope Si-NCs started  about 20 years ago. The results obtained  revealed that doped Si-NCs showed different properties with respect to the doped Si-bulk \citep{OlivaNano2016,ArducaMSSP2017}. 

For example a shallow impurity level in bulk may become a deep level at the nanoscale. In bulk Si the level of doping lies in the interval 10$^{13}$-10$^{18}$ cm$^{-3}$. In a Si-NC of about 2 nm in diameter, which possess more than 200 Si atoms, the introduction of a single impurity atom corresponds to a doping level of about 10$^{20}$ cm$^{-3}$, that is a quite different situation. Despite the large number of works dedicated to the study of doped and codoped Si-NCs, important issues still exist. The effective dopant location (within the nanocrystals, at their surface-interface, or in the surrounding matrix?), the occurrence of self-purifications effects (that is the tendency of the Si-NC to expel the dopant atoms to its surface) as well as the role played by the chemical environmental of the Si-NC are 
issues not yet solved by  the scientific community \citep{PolisskiPB1999,OlivaNano2016,ArducaMSSP2017, MelnikovPRL2004,CantelePRB2005,OssiciniAPL2005,DalpianPRL2006,ChanNL2008,
NorrisScience2008,LiPRB2008,ChelikowskyRPP2011}. 

Experimentally, several factors contribute to make the interpretation of the measurements on these
systems a difficult task. First of all, independently on the fabrication technique, in experimental samples there aren't two identical Si-NCs. For instance, samples show a strong dispersion in the Si-NC size, that is difficult to determine. In this case it is possible that
the observed quantity does not correspond exactly to the mean size
but instead to the most responsive Si-NCs \citep{CredoAPL1999}. Moreover,
Si-NCs synthesized by different techniques often show different
properties in size, shape and in the interface structure. Finally, in
solid nanocrystal arrays some collective effects caused by electron,
photon, and phonon transfer between Si-NCs can render more complicated the interpretation of the experimental results. 
 
The aim of this review is to summarize the latest progress in the fascinating research field related to the understanding of the doping in Si-NCs with n- and p-type impurities. It should be pointed out that in these last few years some review articles on this topic have been published\citep{MangoliniJVST2013,OlivaNano2016,ArducaMSSP2017,PereiraJPD2015,NiJPD2015}. At variance with these reviews, since the role of theoretical modelings and simulations has becomed more an more important, we will give particular attention to the theoretical outcomes.  Moreover we will focus our attention not only on single doped Si-NCs but also on compensated codoped Si-NCs, a very interesting and rapidly growing field. Mainly, we will consider B and P as dopant atoms due to their high solid solubility in silicon and because they are the most commonly used impurities in experiments.

In this work we will use the notation "doped Si-NCs" to indicate systems where B and/or P atoms are incorporated in  Si-NCs. However, in principle, it would be better to speak about impurity atoms in the case of simple incorporation and of dopants in the case of activated impurities. Nevertheless in line with the use in the literature we will adopt the term dopant; it will be clear from the discussion the real meaning of this term. 

The review is organized as follows. First, in section \ref{sec:theo} we will sketch the ab-initio models used for the calculations of the structural, electronic and optical properties of the nanostructures. In section \ref{sec:free} we will review the experimental results regarding doped free-standing Si-NCs, with particular focus on the theoretical outcomes (see subsections \ref{sub:formen} and \ref{sub:ioniz}). Section \ref{sec:matr} is instead devoted to the discussion of the properties of matrix-embedded doped Si-NCs; a subsection is dedicated to the theoretical results (subsection  \ref{sub:formen1}). Transport is discussed in section \ref{sec:trans}, whereas section \ref{sec:codo} is entirely devoted to the study of codoped compensated Si-NCs. Conclusions are presented in section \ref{sec:concl}.

\section{Theoretical Modelling}
\label{sec:theo}
 
\begin{figure}
\begin{center}
\includegraphics[width=7.41cm,height=7.36cm]{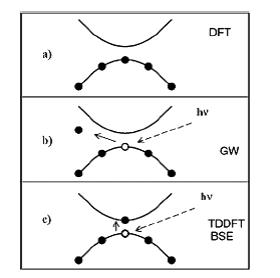}
\caption{(Color online) Scheme of the three processes described in the text: (a) ground state, described by DFT,(b) photoemission experiment (charged excitation) described by GW, (c) absorption experiment (neutral excitation) described by Bethe-Salpeter equation and/or by  Time Dependent DFT. Reprinted with permission from Ref. \citep{DegoliCR2009}}.
\label{fig:figure3}
\end{center}
\end{figure}

The structural, electronic and optical properties of complex systems are nowadays accessible, thanks to the impressive development of theoretical approaches and to the availability of High-Performance Computing platforms. Surfaces, nanostructures, and even biological systems can now be studied within ab-initio methods \citep{DegoliCR2009,ChelikowskyRPP2011}.  
In principle, within the Born-Oppenheimer approximation to decouple the ionic and electronic dynamic, the equation that governs systems of strongly interacting particles is the Many-Body (MB) equation whose solution is a formidable task.
Among all, the single-particle Density Functional Theory (DFT) is one of the most popular methods to study ground state properties of systems of interacting particles. This approach allows to map an interacting quantum MB system into a fictitious system of non-interacting particles.
The Green function approach, instead, permits to calculate the excitations of a system of  interacting particles by mapping the MB electronic problem into a problem of weakly interacting quasi-particles, where the quasi-particle describes the particle plus its screened interaction with the rest of the system.
Fig.~\ref{fig:figure3} shows some schemes used to calculate ground state properties, band structures, and optical spectra: DFT for ground state properties, GW for band structure calculations (charged excitations), and the Bethe-Salpeter equation (BSE) and the time dependent (TD) DFT approaches for optical spectra (neutral excitations). 

Concerning DFT, it is worth noting that the Kohn-Sham (KS) equations, on which the method is based, represent a fictitious auxiliary system with no physical meaning. Nevertheless, their eigenvalues are often interpreted as one-electron excitation energies corresponding to the excitation spectra of the system upon removal or addition of an electron, and DFT is in this way used to calculate band structures. The qualitative agreement with experiments is often remarkable. The quantitative agreement, instead, depends on the approximations used to describe the exchange-correlation energy in the KS equations. The most used approximations, named local density approximation (LDA) and generalised gradient approximation (GGA), lead to a systematic underestimation of the semiconductor energy gap.
This fact can be overcomed by computing quasiparticle energies within the GW scheme. This approach represents the most refined method for the band structure calculation of solids, surfaces, nanostructures  and for the determination of the energy levels of molecules and atoms.
GW well describes direct and inverse photoemission spectroscopies (see panel b of Fig.~\ref{fig:figure3}), where the final state is a charged system since one electron have been removed or add to the system, but fails to describe optical excitations. A correct description of optical spectra (see panel c of Fig.~\ref{fig:figure3}) for both free-standing and embedded Si-NCs requires an appropriate treatment of the local field effects \citep{GuerraPSSB2010,GuerraPRB2011} and an exhaustive calculation of the electron-hole interaction, that can be obtained by solving the BSE within the Many-Body Perturbation Theory (MBPT) \citep{OnidaREVMODPHYS2002}.

In the DFT+MBPT approach, therefore, a detailed estimation of the optoelectronic properties of a system requires: (i) the optimization of the atomic positions  obtained by minimizing the interatomic forces on each atom without any symmetry constraint, (ii) the calculation of the Kohn-Sham states and energies, (iii) the inclusion of quasiparticle corrections withing the GW approximation and (iv) the resolution of the excitonic Hamiltonian, for instance using the BSE approach.

It is now a widespread belief among the nanoscience research community that ab-initio
approaches constitute a unique and very powerful instrument to control and design the properties of novel materials and devices with an accuracy that complements experimental observations \citep{OnidaREVMODPHYS2002}.
 
In the next sections we will review, together with experimental outcomes, theoretical results regarding both free-standing (Section \ref{sec:free}) and matrix-embedded doped Si-NCs (Section \ref{sec:matr}) with particular attention to the determination of formation energies and  self-purification effects, the electronic and optical properties, the role of the Si-NC surfaces and interface terminations.  Most of the presented results habe been obtained  within the DFT  using the LDA for the exchange-correlation energy. As previously pointed out, this approximation leads to an underestimation of semiconductor energy gaps. 
Nevertheless the results obtained through LDA are interesting for several reasons: i) the inclusion of MB effects is very computationally demanding and thus limited to very small nanocrystals \cite{IoriPRB2007}, ii) an almost complete compensation of self-energy and excitonic effects \cite{GuerraPRB2009,LuppiPRB2005,GuerraSM2009} has been observed in silicon based nanostructures, thus rendering the LDA based optoelectronic outcomes useful, iii) often scientists are interested in trends and trends will remain similar on going from the independent-particle approximation to the many-body approach. 

\section{Free-standing Nanocrystals}
 \label{sec:free}
Free-standing Si-NCs are attractive due to the possible fabrication of large area optoelectronic devices by vacuum-free printable processes  and for their use in biological applications. Researchers have employed several methods to obtain free-standing Si-NCs, among these we note laser pirolysis\citep{EhbrechtPRB1999} and thermal dissociation
\citep{WilsonScience1993,OnischukJAS1997,OstraatAPL2001}, laser ablation\citep{WerwaAPL1994}, plasma synthesis\citep{MangoliniNL2005}, liquid \citep{HolmesJACS2001} and solid phase synthesis\citep{HesselCM2012}. Several excellent reviews about different routes utilized to synthetize free-standing Si-NCs are appeared\citep{MangoliniJVST2013,KnippingJNN2004,VeinotCC2006,KortshagenCR2016,YinNAT2005}.   
\begin{figure}
\begin{center}
\includegraphics[width=7.74cm,height=6.5cm]{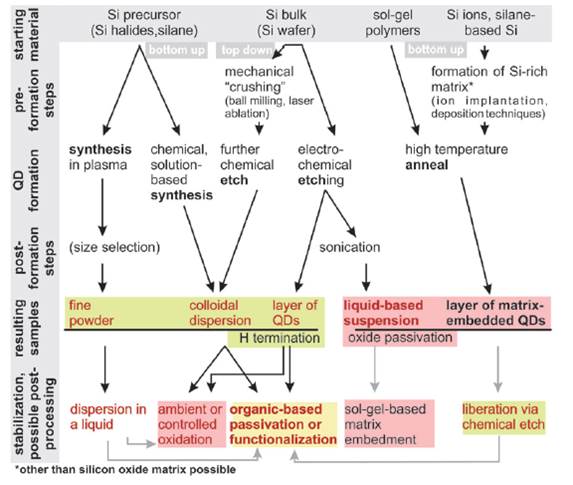}
\caption{(Color online) A general overview of the most common sequences of steps for generating free-standing and matrix-embedded Si-NCs. Free-standing cases are shown in red. Reprinted with permission from Ref. \citep{GregorJPCM2014}.}
\label{fig:figure0}
\end{center}
\end{figure}
Figure \ref{fig:figure0} shows an overview of the steps usually followed for the generation of free-standing Si-NCs. Through these techniques is possible to have a good control of the Si-NCs size and shape \citep{VeinotCC2006} and to synthetize high quality  Si-NCs \citep{PereiraPRB2012} with PL peaks spanning the entire visible spectrum \citep{PiNANO2008,DasogACS2014} and with a high quantum yield up to 60$\%$ \citep{JurbergsAPL2006}.
The surface of the Si-NCs can be covered by different species and functional groups \citep{GregorJPCM2014,VeinotCC2006}. Dasog et al.\citep{DasogACS2014}, for instance, have synthetized colloidal Si-NCs of 3-4 nm in diameter functionalized with alkyl, amine, phosphine, and acetal functional groups. They proved that the PL of the Si-NCs can be effectively tuned across the entire visible spectral region without changing particle size, thus elucidating the role of surface termination. 
Doped Si-NCs with diameter down to 3 nm have been synthetized using different strategies \citep{PereiraJPD2015,NiJPD2015,PiAPL2008,BaldwinCC2006,SomenoJAP2012,MaJNR2012}. Among all, Fujii and coworkers developed an efficient procedure \cite{SugimotoJAP2011,SugimotoJPCC2012,SugimotoJPCC2013a,SugimotoJPCC2013b,FujiiJAP2014,KannoJAP2016,AlmeidaPRB2016} where
Si, SiO$_2$, P$_2$O$_5$ or B$_2$O$_3$ were simultaneously sputter-deposited and then annealed in N$_2$ gas atmosphere. During annealing the B- or P-doped Si-NCs were grown in borosilicate (BSG) or phosphosilicate (PSG), i. e. B- or P-doped silica, glass matrices. Then the Si-rich BSG or PSG films were peeled from the plates and annealed in a N$_2$ gas atmosphere to obtain Si-NCs embedded in BSG or PSG matrices. To isolate Si-NCs from the matrices, these films were dissolved in HF solution in ultrasonic bath. This process led to a HF solution containing isolated Si-NCs, that were separated from the solution by centrifugation obtaining in the concentrator a Si-NC powder. Methanol was then added to disperse the Si-NCs.

The properties of free-standing doped Si-NCs have been investigated by absorption, PL and PL excitation (PLE) spectroscopies
\citep{DasogACS2014,PiAPL2008,BaldwinCC2006,SugimotoJAP2011,SugimotoJPCC2013a,SalivatiSS2011,ZhouAPL2014,ZhouPPSC2015,KramerNL2015,ZhouACS2015,NiAOM2016,ZhouACS2016},
atomic force microscopy (AFM) \cite{XuJAPP2014}, atom probe tomography (APT) and proximity histogram analysis \citep{NomotoJPCC2016}, electrically detected magnetic resonance (EDMR) \citep{AlmeidaPRB2016,PereiraPRB2009,NakamineJJAP2011}, electron paramagnetic resonance (EPR) \citep{BaldwinCC2006,AlmeidaPRB2016,KramerNL2015,PereiraPRB2009,PereiraPHYS2007,LechnerJAP2008,StegnerPRL2008,StegnerPRB2009}, field effect transistor (FET) measurements \citep{GresbackACS2014}, Fourier transform infrared spectroscopy (FTIR) \citep{KramerNL2015,ZhouACS2015,NiAOM2016,ZhouACS2016}, Kelvin probe force microscopy\cite{XuJAPP2014}, nuclear magnetic resonance (NMR) \citep{BaldwinCC2006}, Raman spectroscopy \citep{FujiiJAP2014,ZhouPPSC2015,LechnerJAP2008}, secondary-ion mass spectroscopy (SIMS) and inductively coupled plasma atomic emission spectroscopy (ICP-AES) \citep{PiAPL2008,StegnerPRB2009}, trasmission electron microscopy (TEM) and high-resolution TEM (HRTEM) \citep{BaldwinCC2006,SugimotoJAP2011,SugimotoJPCC2013a,AlmeidaPRB2016,ZhouPPSC2015,ZhouACS2015,ZhouACS2016} and, finally, x-ray photoelectron spectroscopy (XPS) \citep{DasogACS2014,SugimotoJPCC2013a,FujiiJAP2014,ZhouPPSC2015,KramerNL2015}. 
\begin{figure}
\begin{center}
\includegraphics[width=10.0cm,height=6.99cm]{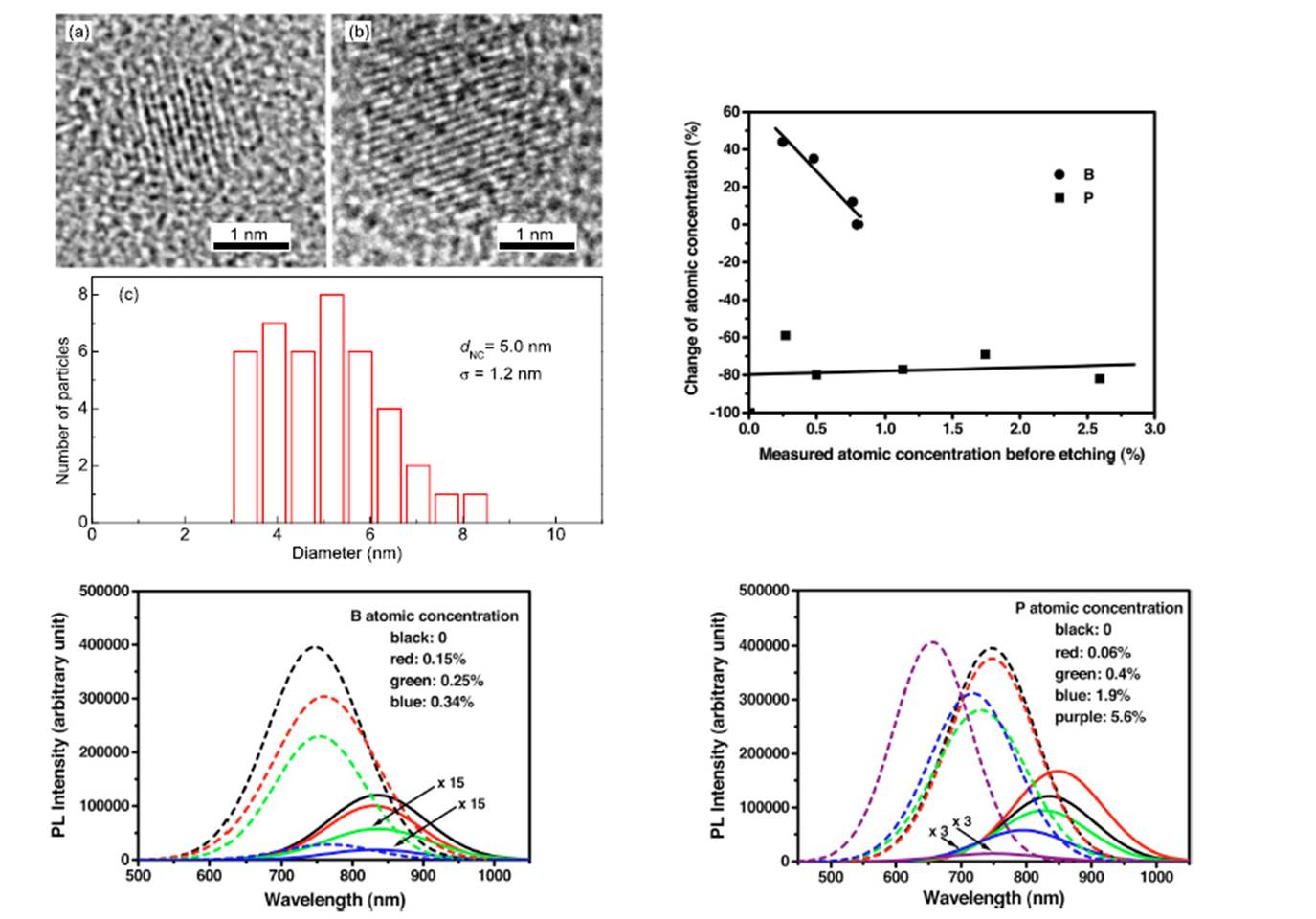}
\caption{(Color online) Top Left Panel: (a) and (b) HRTEM images of P-doped Si-NCs terminated by hydrogens. The lattice fringes correspond to the $\{111\}$ planes of Si. (c) Si-NCs size distribution extracted from the HRTEM images. Reprinted with permission from Ref. \citep{AlmeidaPRB2016}. Top Right Panel: change of dopant concentration after HF etching for B- and P- doped Si-NCs that have been exposed to air for 5 days. Solid lines result from linear fits to the data. Reprinted with permission from Ref. \citep{PiAPL2008}. Bottom Left Panel: PL spectra from as-synthesized intrinsic B-doped Si-NCs (solid lines) and the same Si-NCs after five-days exposure to air at room temperature (dashed lines). The B-doped Si-NCs are labeled according to the B atomic concentrations. The intensity of PL from as-synthetised Si-NCs with B concentrations of 0.25 $\%$ and 0.34 $\%$ is magnified by a factor 15. Reprinted with permission from Ref. \citep{PiAPL2008}. Bottom Right Panel: PL spectra from as-synthesized intrinsic P-doped Si-NCs (solid lines) and the same Si-NCs after five-days exposure to air at room temperature (dashed lines). The P-doped Si-NCs are labeled according to the P atomic concentrations. The intensity of PL from as-synthetised Si-NCs with P concentrations of 1.9 $\%$ and 5.6 $\%$ is magnified by a factor 3. Reprinted with permission from Ref. \citep{PiAPL2008} }
\label{fig:figureexp1}
\end{center}
\end{figure}
\begin{figure}
\begin{center}
\includegraphics[width=7.74cm,height=6.99cm]{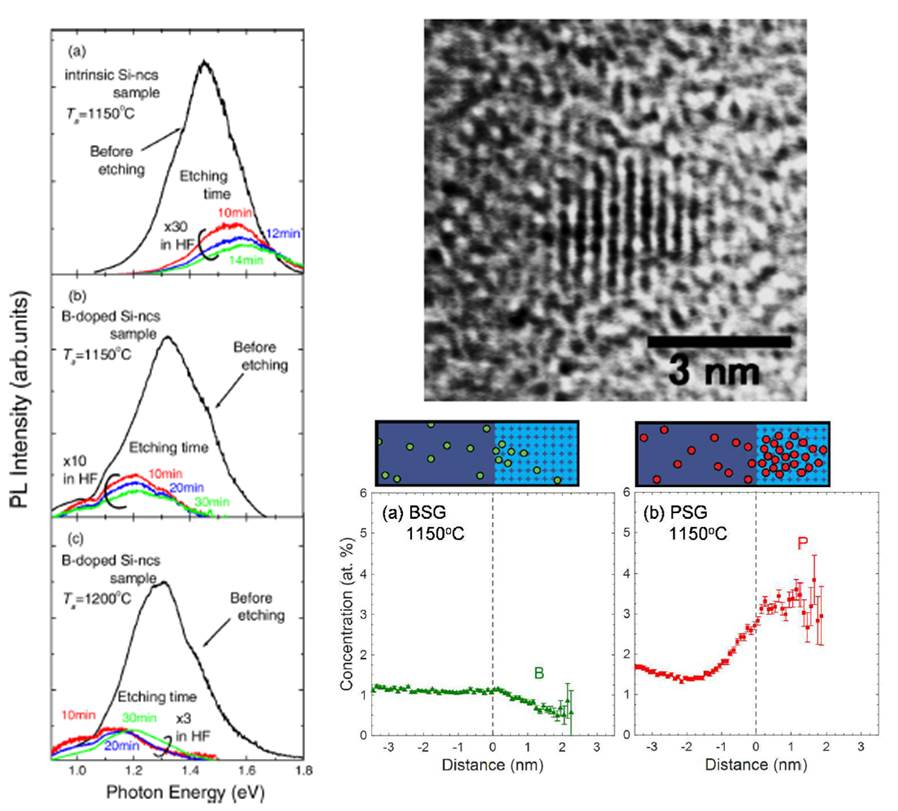}
\caption{(Color online) Left: PL spectra taken before and during HF etching. The etching time is indicated in the graph. (a) Intrinsic Si-NCs samples annealed at 1150 $\degree$C, (b) B-doped Si-NCs samples annelaed at 1150 $\degree$C, and (c) B-doped Si-NCs samples annealed at 1200 $\degree$C. Reprinted with permission from Ref. \citep{SugimotoJAP2011}. Right Top: HRTEM image of B-doped Si-NCs after HF etching.  Reprinted with permission from Ref. \citep{SugimotoJAP2011}. Right Bottom: proxigram analysis of B- and P- doped Si-NCs in (a) borosilicate glass and (b) phosphosilicate glass annealed at 1150 $\degree$C. On top of each figure the model images of the B or P distribution. B atoms are green balls and P atoms red balls. Reprinted with permission from Ref. \citep{NomotoJPCC2016}.}
\label{fig:figureexp2}
\end{center}
\end{figure}

Resuming the experimental information (see Fig. \ref{fig:figureexp1} and Fig. \ref{fig:figureexp2}) one can say that the dopant concentration can be adjusted in a very wide range \citep{PiAPL2008,ZhouPPSC2015,LechnerJAP2008,StegnerPRB2009} even if only a small part of the dopants are activated \citep{StegnerPRB2009,GresbackACS2014,RoweNL2013}. Moreover doping efficiency depends both on Si-NCs size and on Si-NCs environment \citep{AlmeidaPRB2016}. For a review on the experimental results regarding doping efficiency and electrical activity of doped Si-NCs the reader is referred to Refs: \citep{PereiraJPD2015,NiJPD2015}. The doped Si-NCs have a good crystallinity, showing that the lattice spacing with respect to undoped Si-NCs is not affected for P-doped Si-NCs, whereas it is slightly reduced for B-doped ones, owing to the smaller size of the B atom \citep{SugimotoJAP2011,ZhouPPSC2015} (see top left panel of Fig. \ref{fig:figureexp1} and bottom panel of Fig. \ref{fig:figureexp2})  

Experiments pointed out that the hyperfine splitting (HFS) shows a clear size dependence \citep{PereiraPRB2009} and that impurities are mainly incorporated in substitutional sites \citep{SugimotoJAP2011,AlmeidaPRB2016,NakamineJJAP2011,PereiraPHYS2007,StegnerPRL2008}. 
Still regarding the dopant location, several studied pointed out that B atoms prefer to stay in the Si-NC core, whereas P atoms prefer to reside at the Si-NC surface \citep{PiAPL2008,ZhouAPL2014,StegnerPRB2009} (see top right panel of Fig. \ref{fig:figureexp1}), even if this result can be reversed depending on the Si-NCs preparation condition \citep{,SugimotoJAP2011,NomotoJPCC2016} (see top right panel of Fig. \ref{fig:figureexp2}). The PL properties of both B-and P- doped Si-NCs show a shift of the PL maximum to lower energies. Regarding the PL intensity, in the case of B doping there is a decrease with an increase in the doping concentration, whereas for P-doping the intensity becomes smaller at small P concentration and subsequently increases at large concentration \citep{PiAPL2008,SugimotoJAP2011} (see the bottom panel  of Fig. \ref{fig:figureexp1} and the left panel of Fig. \ref{fig:figureexp2}).      

\subsection{Ab-Initio calculations: formation energies, role of sizes, surfaces, capping species and vacancies}
\label{sub:formen}
From a theoretical point of view, DFT-based ab-initio calculations for free-standing Si-NCs have been performed in real space 
\citep{ChelikowskyRPP2011} or in reciprocal space \citep{DegoliAQC2009}. The first approach doesn't show problems connected with spurious interaction between replicas of NCs. The second one, implementing a  supercell approach, requires large cells (thus large vacuum regions) to prevent interactions between a NC and its image. Moreover in the case of charged systems, the reciprocal space approach requires a procedure to neutralise the background charge.
The dangling bonds at the surface of the Si-NCs are usually passivated with H atoms,
which excludes the occurrence of effects induced by the presence of surface dangling bond states and surface reconstructions. Although Si-NCs are usually oxidized, there are numerous experiments where
the dangling bonds at the surface are passivated with H \citep{Fujiibook2010}. Moreover, as pointed out by several authors, in H-terminated Si-NCs HOMO and LUMO wavefunctions don't contain contributions from H atoms \citep{LuppiJAP2003,ZhouNL2003,LuppiPRB2005,ZhouJACS2003,DegoliPRB2004,RamosPRB2005}. As a consequence, changing in the energy gap can be only induced by size reduction (QC effect) or by introducing substitutional impurities.
\begin{figure}
\begin{center}
\includegraphics[width=12cm,height=7.98cm]{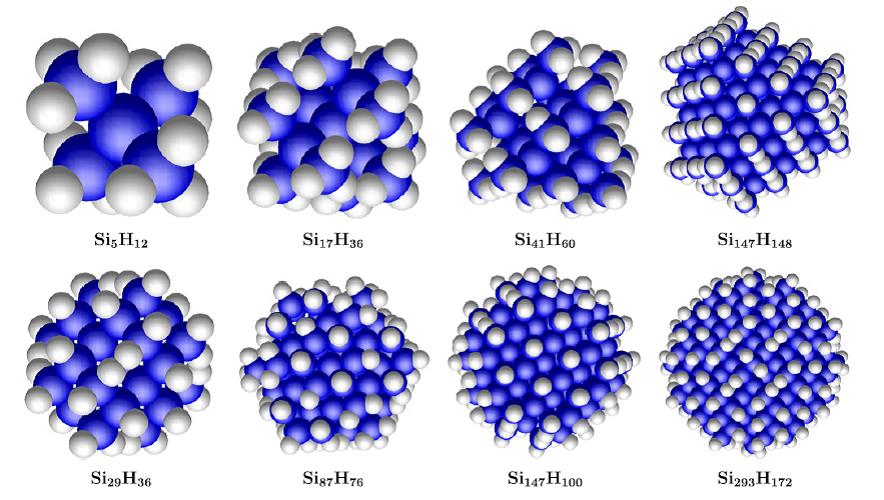}
\caption{(Color online) Faceted (top) and spherical (bottom) undoped Si-NCs. The dark spheres represent the Si atoms and the light spheres represent the H atoms. The faceted Si-NCs and their average diameters d calculated  as two times the average distance of the H atoms with respect to the center of the Si-NC  are the Si$_5$H$_{12}$ (d = 0.64 nm), the Si$_{17}$H$_{36}$ (d = 0.94 nm), the Si$_{41}$H$_{60}$ (d = 1.26 nm), and the Si$_{147}$H$_{148}$ (d = 1.88 nm). The spherical Si-NCs and their average diameters d are the Si$_{29}$H$_{36}$ (d = 1.08 nm), the Si$_{87}$H$_{76}$ (d = 1.56 nm), the Si$_{147}$H$_{100}$  (d = 1.86 nm), and the Si$_{292}$H$_{172}$ (d = 3.0 nm). Due to its small size the Si$_{5}$H$_{12}$-NC could be classified either as spherical-like or as faceted. The atomic sites of the Si-NCs at their initial configuration resemble locally the ones of the diamond structure (T$_d$ point symmetry), where the atoms have a tetrahedral coordination. In both spherical-like and faceted Si-NCs, there is a Si atom at the center. Reprinted with permission from Ref. \citep{RamosJPCM2007}.}
\label{fig:figure4}
\end{center}
\end{figure}

Concerning the shape of the Si-NCs, one can consider both spherical and faceted Si-NCs. The spherical Si-NCs are obtained, from Si bulk, by cutting Si atoms outside a sphere of a particular diameter, whereas the faceted Si-NCs result from a shell-by-shell construction procedure, in which one starts from a central atom and adds shells of atoms successively. Figure~\ref{fig:figure4} shows a cartoon of faceted and spherical Si-NCs. 

In the Si-bulk crystal B and P impurities replace substitutionally Si in the lattice and show sp$^3$ hybridization, while in Si-NCs the dopant atom could be either inside or at the surface. Therefore in the case of Si-NCs the considered dopant locations are all the substitutional sites, starting from the center of the Si-NC and moving the impurity to the other substitutional sites in the entire Si-NC (one follows this procedure because both EPR \citep{StegnerPRL2008} and EDMR \citep{NakamineJJAP2011} have demonstrated that impurities are not interstitials).
By taking into account only one impurity per Si-NC one can simulate a wide range of high doping concentrations, which vary approximately from 2x10$^{20}$ cm$^{−3}$ (1 mol$\%$) to 7x10$^{21}$ cm$^{−3}$ (20 mol$\%$).

Starting from a Si$_n$H$_m$-NC the formation energy for a neutral X impurity can be defined as the energy
needed to insert the X atom with chemical potential $\mu_X$ within the Si-NC after removing a Si atom, that is:

\begin{equation}
E_f = E(Si_{n-1} X H_m) - E(Si_n H_m) + \mu_{Si} - \mu_X
\label{eq1}
\end{equation}

where E is the total energy of the system, $\mu_{Si}$ is the total energy per atom of bulk Si, and $\mu_X$ is the total energy per atom of the impurity.

Before looking at the results of the formation energy, it is interesting to look at the changes induced on the structural properties by the presence of an impurity placed at the centre of the Si-NC 
\citep{MelnikovPRL2004,MelnikovPRL2004,IoriPRB2007,RamosJPCM2007,ZhouPRB2005,ZengJN2012}. 
After ionic relaxation performed without symmetry constraints, the first neighbors of the impurity in the Si-NC experience different displacements. For the B-doped Si-NCs, while the Si-Si bond lengths remain almost unchanged, some reconstructions occur around the impurity. The overall structure has $C_{3v}$ symmetry, with an impurity displacement along the $<111>$ direction. Such displacement leads to one longer and three shorter and equal Si-impurity distances. While the
longer bond is “almost” independent on the Si-NC size, the shorter one increases with the Si-NC size. It is interesting to note that, instead, the relaxation of a bulk Si supercell containing the B impurity
leads to an “almost” $T_d$ configuration, in which the four B-Si bonds are practically the same. The relaxation of the P-doped Si-NCs leads to a nearly $T_d$ symmetry, in which the differences between the four P-Si bonds are negligible. This means that, for doped Si-NCs, there is a difference in the behavior between trivalent and pentavalent atoms, i.e. the amount of the relaxation around the impurity is directly related to the impurity valence. Concerning the dependence on the shape of the Si-NCs,  the differences of bond lengths are clear. As a rule, the Si-X bond lengths of faceted Si-NCs are longer than the ones calculated for the spherical ones, even when they have similar average radii. 

\begin{figure}
\begin{center}
\includegraphics[width=12.cm,height=11.88cm]{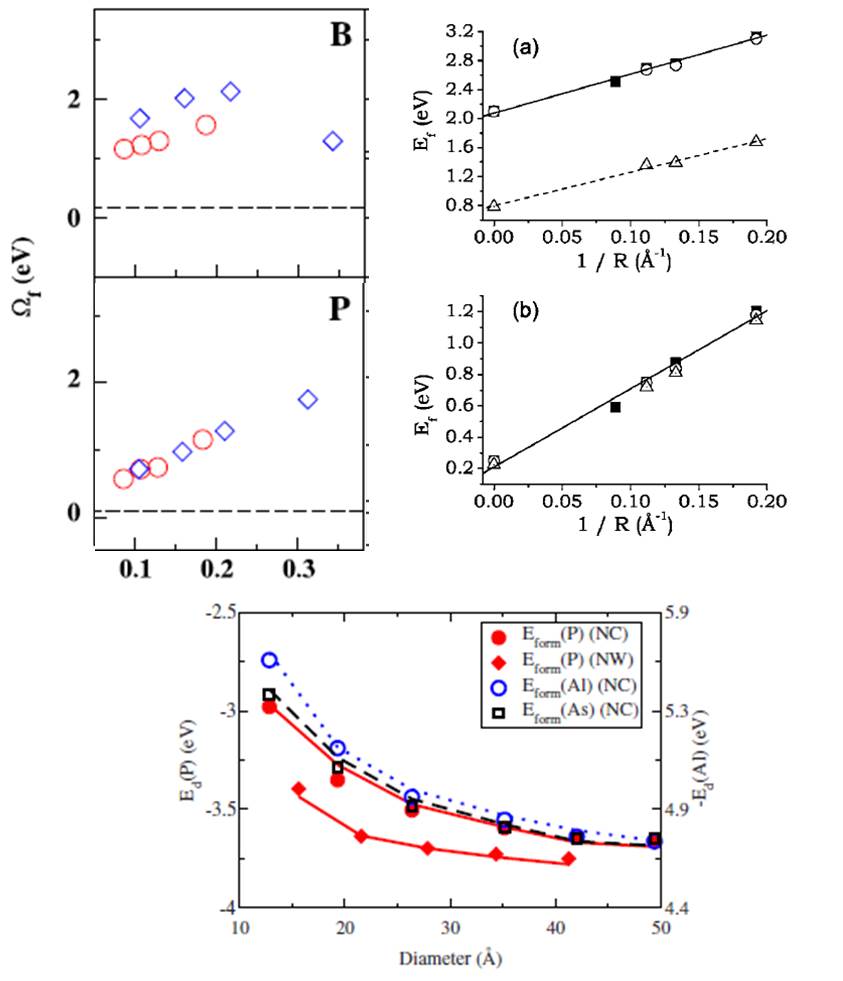}
\caption{(Color online) Top Left Panel: impurity formation energies of B and P impurities located at the Si-NC center versus the reciprocal radii of the doped Si-NC. The energy values are given for faceted (diamonds) and spherical (circles) Si-NCs. The dashed lines indicate the impurity formation energies in Si bulk. Reprinted with permission from Ref. \citep{RamosJPCM2007}.
Bottom Panel: variation in the Kohn-Sham eigenvalues of the defect (impurity) levels E$_d$ (lines) and the formation energy E$_{form}$ (points) as a function of the nanocrystal (NC) or nanowire (NW) diameter. The solid and dashed lines correspond to E$_d$(P) and E$_d$(As), respectively.  The points represent the formation energies. E$_{form}$ for each dopant are shifted uniformly to align them with their
corresponding E$_d$ curve. Reprinted with permission from Ref. \citep{ChanNL2008}. Top Right Panel : formation energy of the neutral impurities at the Si-NC center, as a function of the inverse Si-NC radius. Both (a) B- and (b) P-doped Si-NCs are considered. The three sets of data correspond to total energies calculated for three different geometries: (i) both doped and undoped Si-NCs have not been relaxed, filled squares; (ii) only the undoped Si-NC has been relaxed, empty circles; (iii) both Si-NCs
have been relaxed, empty triangles. The zero inverse radius corresponds to bulk Si. The lines are linear fits of the corresponding set of data. Reprinted with permission from Ref. \citep{CantelePRB2005}.}
\label{fig:figure5}
\end{center}
\end{figure}

Formation energies, calculated from  Eq. \ref{eq1} and reported in Fig.~\ref{fig:figure5}, are strongly connected with the structural properties of the system. For B and P, the formation energy decreases as the Si-NC size increases, thus showing that larger Si-NCs can more easily sustain the B- or P-doping.
It also shows an almost linear dependence on the reciprocal radius of the Si-NCs. A comparison between faceted and spherical Si-NCs shows that the incorporation of a B impurity costs more energy for the faceted Si-NC, whereas is nearly shape-independent for P. For both B- and P-doping the impurity formation energies are always higher in Si-NCs than in Si bulk. Moreover, for similar radius, calculations pointed out that B incorporation cost more than P segregation, in agreement with experimental outcomes \cite{PiAPL2008}. Similar results have been obtained by other authors, confirming that the effect of size is much more evident for B-doping that for P-doping.   \citep{MelnikovPRL2004,ChanNL2008,ZengJN2012,OssiciniJSTQE2006,OssiciniSS2007,MagriJCMSE2007,DalpianPRLR2008,
MaPRB2013}. 

It was also possible to bring out the role of the relaxation with respect to that of Si-NC size
(see Fig.~\ref{fig:figure5}). Three different sets of data were presented and discussed \citep{CantelePRB2005}.
In the first situation (filled squares) the total energies were calculated keeping the Si-Si and Si-impurity distances fixed at the Si-Si bond length of bulk Si. In the second case (empty circles) the undoped Si-NC was firstly relaxed, and the optimized geometry was used for the doped Si-NC, thus neglecting the effects of the relaxation around the impurity. In the third situation (empty triangles) the doped Si-NC was fully relaxed after the insertion of the impurity. Looking at the results of Fig.~\ref{fig:figure5}, the atomic relaxation around the B impurity induces a significant reduction of the formation energy, which is nearly independent on the nanocrystal size. The same reduction is observed for P-doped Si-NCs, even if it is much less evident. Moreover, for impurity atom located at the Si-NC center, the variation of the formation energy with 1/R cannot be attributed only to the atomic relaxation around the impurity. This behavior is strongly connected with the QC effect and appears even without any relaxation (see the filled squares).
\begin{figure}
\begin{center}
\includegraphics[width=12.cm,height=8.46cm]{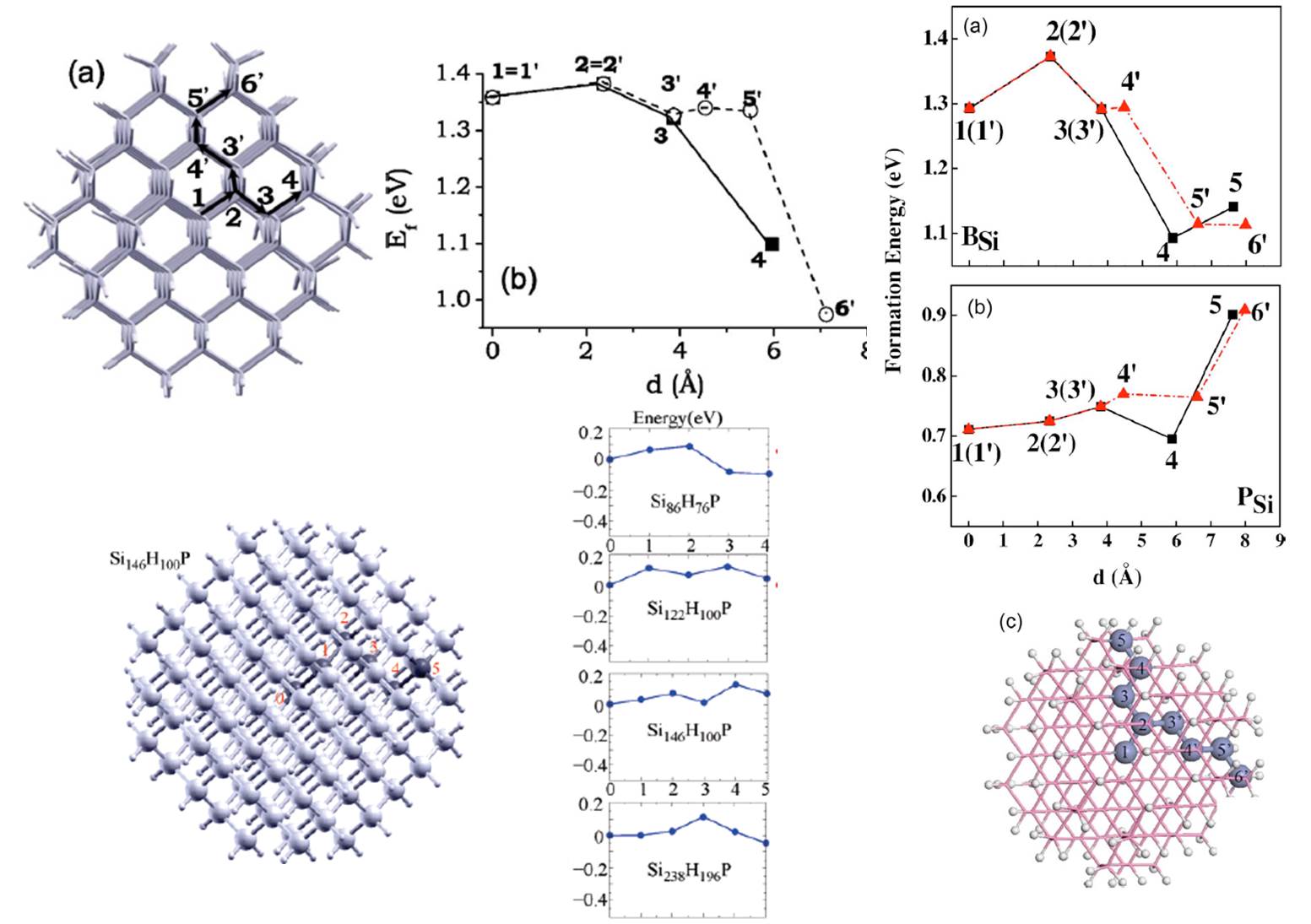}
\caption{(Color online) First and second column, top:
(a) the B impurity is moved along two different paths toward the Si-NC surface. (b) Formation energies for B neutral impurities as a function of the impurity position within the Si-NC. Reprinted with permission from Ref. \citep{CantelePRB2005}. First and second column, bottom, left:  a perspective view of a Si-NC. The lightly shaded atoms correspond to Si atoms, and the small atoms on the surface of the nanocrystal are H atoms. The heavily shaded atoms correspond to the Si atoms that will be substituted by the P impurity. The numbers measure the distance away from the origin in the unit of Si bond length. Right: difference in energy as the P atom moves away from the center of the Si-NC. The energies are with respect to the energy of the Si-NC with P at the center. The x-axis measures the distance of the P atom away from the origin as in left. The numbers measure the distance away from the origin in the unit of Si bond length. Reprinted with permission from Ref. \citep{ChanNL2008}. Third column: the calculated formation energies of (a) B and (b) P impurities at different sites along two different paths in the Si-NC as shown in(c). Reprinted with permission from Ref. \citep{XuPRB2007}}
\label{fig:figure6}
\end{center}
\end{figure}

Important efforts were dedicated to the analysis of the formation energies as a function of the impurity position within the Si-NC \citep{MelnikovPRL2004,CantelePRB2005,OssiciniJSTQE2006,XuPRB2007,ChanAPL2011}. This is an important point because the optical and electrical activities of the dopant atoms depend on their positions.  Figure~(\ref{fig:figure6}) (top) shows the formation energy for the B neutral impurity in a Si$_{146}$BH$_{100}$-NC \citep{CantelePRB2005}. The impurity is moved from the Si-NC center toward the surface along two paths. Since in the experiments 
the Si-NC surfaces depends on the synthesis conditions, impurity atoms directly located  at the Si-NCs surface, i.e. bonded to H atoms, were not considered.  The results show that the B atom tends to migrate toward the subsurface position. This is explained by considering that such positions are the only ones that allow a significant atomic relaxation around the impurity, because in the other cases the surrounding Si cage is quite stable. Thus, as the B atom is moved toward the surface the formation energy decreases, making subsurface positions more stable. The preference of B to be incorporated subsurface has been experimentaly established in the case of B-doped Si-NCs passivated by deuterium \citep{SalivatiSS2011}. The bottom of Figure~(\ref{fig:figure6}) shows the corresponding results for a P impurity \citep{ChanAPL2011}. It is clear that, in this case, for Si-NCs, whose diameter is smaller than about 2 nm, P tends to substitute Si near the surface. Otherwise, the Si-NC center is the energetically most favorable position. Similar results for B and P dopants are shown in the right panel of Fig.~\ref{fig:figure6}  \citep{XuPRB2007}. 

These results have been interpreted in term of a self-purification effect for the smaller Si-NCs: below a diameter of about 2 nm the impurity atoms will be energetically expelled toward the surface and will be hardly incorporated  \citep{MelnikovPRL2004,DalpianPRL2006,CantelePRB2005,ZengJN2012,XuPRB2007,ChanPRB2010}. 

In particular, Chan et al. \citep{ChanNL2008,ChanPRB2010} showed that this effect can be related to both the binding energy (BE) of the dopant (see below) and the interaction of the dopant with the surface. The binding energy contribution always originates a energy minimum at the Si-NC center, while the strain induced by the dopant dominates energy considerations at the surface. The competition between these two effects dictates the presence or not of self-purification effects. 

However, the self-purification effects can be strongly influenced by the presence of defects, vacancies or different capping species used to saturate the dangling bonds at the interface; this fact explains  the efficient incorporation of dopant atoms in small Si-NCs with diameter less than 2 nm \citep{NomotoJPCC2016,PeregoNANO2010,PeregoSIA2013,NomotoMRS2016,NomotoPSS2017}. 

\begin{figure}
\begin{center}
\includegraphics[width=12.cm,height=8.90cm]{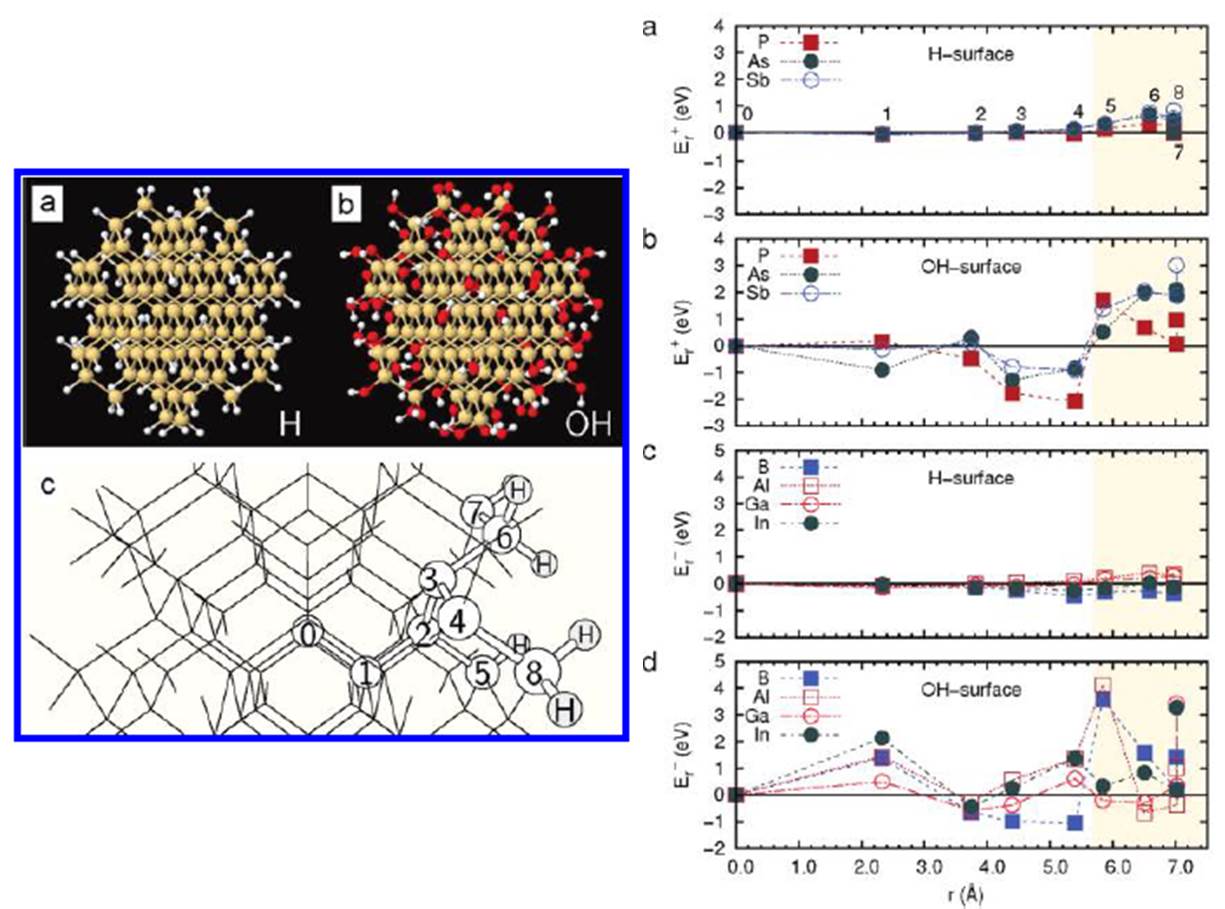}
\caption{(Color online) Left Panel: geometry of (a) Si-NC with H-terminated atoms, and (b) Si-NC with OH terminated atoms. Si, O, and H species are represented by large, medium, and small spheres, respectively. (c) Identification of the substitutional position for the group III and V impurities.  Right Panel: relative energy of (a) ionized group V donors in Si-NCs with H-terminated surface and (b) with OH-terminated surface, (c) ionized group III acceptors in Si-H-NC and (d) in Si-OH-NC, as a function of the average distances to the center of the Si-NC. The points in the shaded area represent surface sites. Reprinted with permission from Ref. 
\citep{CarvalhoJPCC2012}.}
\label{fig:figure7}
\end{center}
\end{figure}

In this context, for example, Chelikosky's group \citep{EomSSC2010} showed that the mechanisms that move the B atom towards the Si-NC surface depend on the presence of a vacancy located at the center of the Si-NC. Vacancies can be formed more easily as the diameter of the Si-NC decreases and, in the presence of a vacancy that relieves the dopant induced stress, the B impurity prefers to stay in the vicinity of the vacancy, thus becoming more stable in the core of the Si-NC and mitigating the self-purification effect.
Ma et al. \citep{MaAPL2011} have studied how imperfections in the passivation of the Si-NC surface influence the dopant formation energy. 
They considered a faceted Si$_{147}$H$_{100}$-NC and they removed a H atom at the surface. The obtained results pointed out that, contrary to the full H-passivated case, the P impurity tends to substitute the Si atom  with the dangling bond at the surface.
On the contrary, the B atom still prefers to stay in a subsurface position, near to the unpassivated Si atoms. In a series of works Pi et al. \citep{PiAPL2008,PiJPCC2011,ChenJPCC2011,PiJnano2012} demonstrated that, for a substitutional P impurity located at the Si-NC surface and possessing different bonding configurations, the most stable structure is obtained when P is not passivated by H while, in the case of B impurity, the most stable configuration is obtained when B is bounded with two H atoms. Moreover  the formations energies for the B-doped Si-NCs are about 1 eV higher than those calculated for the corresponding P-doped cases, showing again that the incorporation of B impurities is less efficient than the incorporation of P impurities.
Finally the role of a different Si-NC termination, namely silanol OH group, has been investigated\citep{CarvalhoJPCC2012,CarvalhoJPCS2011,CarvalhoPSSA2012,CarvalhoPSSB2013,GuerraJACS2014}.  
Carvalho et al. \citep{CarvalhoJPCC2012}, for instance, considered the  OH-terminated Si$_{86}$H$_{76}$-NC (about 1.5 nm of diameter) that was obtained by replacing H capping atoms with OH groups. 
Through the calculation of the dependence of the formation energy on the dopant site, an important difference between hydride- and silanol-terminated Si-NCs emerged (see Fig. \ref{fig:figure7}). As in previous studies \citep{MelnikovPRL2004, CantelePRB2005} their results pointed out that for H-terminated Si-NCs the P impurity prefers to stay in the core of the Si-NC, whereas for the B impurity the subsurface positions are the preferred.  However in the case of OH-terminated Si-NC, the variation of the formation energy with respect to the impurity site occupation is much larger (of the order of 3 eV) than in the case of the Si-NC terminated by H atoms (of the order of 1 eV). Moreover on going from H-terminated to OH-terminated Si-NCs, whereas the B impurity still prefers the subsurface positions, this position become also preferred  by the P impurity. Indeed in both cases the surface positions, where dopants form bonds with oxygen, are in general not the favored ones. 

All these results show how the properties of doped Si-NCs strongly depend not only on the Si-NC size but also on Si-NC surfaces and interfaces.

As discussed before from the experimental point of view it is difficult to obtain information about the exact dopant location, nevertheless different techniques have been employed to this task. Regarding as-grown doped Si-NCs sequentially oxidized and HF etched \cite{PiAPL2008} it seems that, in contrast with the theoretical results for H-terminated Si-NCs, B prefers to stay in the Si-NC core whereas P prefers to stay at the Si-NC surface. Nevertheless theory has proven that, for what concern the H-terminated Si-NCs with H deficiencies, the P impurity prefers to be located at the Si-NC surface and B in the core \cite{MaAPL2011}, in agreement with the experiments. Moreover, it is well know that experimental results depend on the Si-NCs synthesis procedures. This was confirmed, for instance, by Sugimoto et al.\cite{SugimotoJAP2011} which demonstrated that, by changing the etching time, it is possible obtain Si-NCs  with an intrinsic core and a heavily B-doped shell.
Moreover, as underlined by several authors \cite{ArducaMSSP2017,ZhouPPSC2015,MaAPL2011}, whereas the theoretical results concern the thermodynamic stability of the impurity atoms in the Si-NCs, in the experiments is not possible to separate the equilibrium properties from the kinetic effects, which play an important role with respect to doping.

Finally a systematic study of the plasmonic properties of Si-NCs doped with B and P through localized surface plasmon resonance demonstrated that the dominant doping mechanism for P is substitutional doping, whereas for B is surface doping \cite{KramerNL2015}.

\subsection{Ab-initio calculations. Activation energies, impurity states, optical properties}
\label{sub:ioniz}

As in the case of bulk Si, doping of Si-NCs is performed in order to tune their electronic properties. However, the simple incorporation of B or P impurities in the Si-NCs does not ensure the generation of free majority charge carriers by ionization of the dopant atom at room temperature. In a Si bulk system the extra hole associated to the B impurity creates an acceptor level, whereas for P impurity the extra electron creates a donor state. The activation of the impurities, of these states, is related to their binding energy (BE). The BE of the acceptor within a bulk crystal is defined as the energy difference between the defect level and the valence band maximum (VBM), whereas the BE of the donor is given by the energy difference between the conduction band minimum (CBM) and the defect level. In Si-NCs, as stated in Section \ref{sec:intro}, as consequence of QC effect, size reduction results in a enlargment of the band gap, consequently after doping the relative position of the defect energy level with respect to VBM and CBM strongly depends on the Si-NC size \citep{XuPRB2007}. Moreover, differently from the bulk, the extra hole or electron tends to remain localized near the impurity\citep{MelnikovPRL2004,OssiciniAPL2005}.

A more precise definition of the BE at the nanoscale is given by the impurity activation energy 
$E_{act}$ \citep{MelnikovPRL2004,CantelePRB2005,ZhouPRB2005,LannooPRL1995} 

\begin{equation}
E_{act}(B) = I_u - A_d
\end{equation}
\begin{equation}
E_{act}(P)= I_d - A_u
\end{equation}

where $I$ and $A$ give the Si-NC ionization energy and electron affinity, respectively,
while the subscripts $u$ and $d$ refer to the pure (undoped) and doped system, respectively.
The activation energy represents the energy required to ionize the P-doped Si-NC by removing an electron minus the energy gained by adding the electron to a pure Si-NC. Therefore, the ionization energy is given by

\begin{equation}
I_d = E(n-1) - E(n)
\end{equation}

where $n$ is the number of electrons in a neutral P-doped Si-NC and $E(n)$ and $E(n-1)$ are the total energies of the ground state for the 
neutral and positively charged P-doped Si-NC.  The electron affinity of undoped Si-NCs is given by

\begin{equation}
A_u = E(n-1) - E(n)
\end{equation}

where $E(n-1)$ and $E(n$) are the ground state total energies of a neutral and negatively charged Si-NC, respectively. Similar equations can be derived for the B impurity.

\begin{figure}
\begin{center}
\includegraphics[width=12.cm,height=8.69cm]{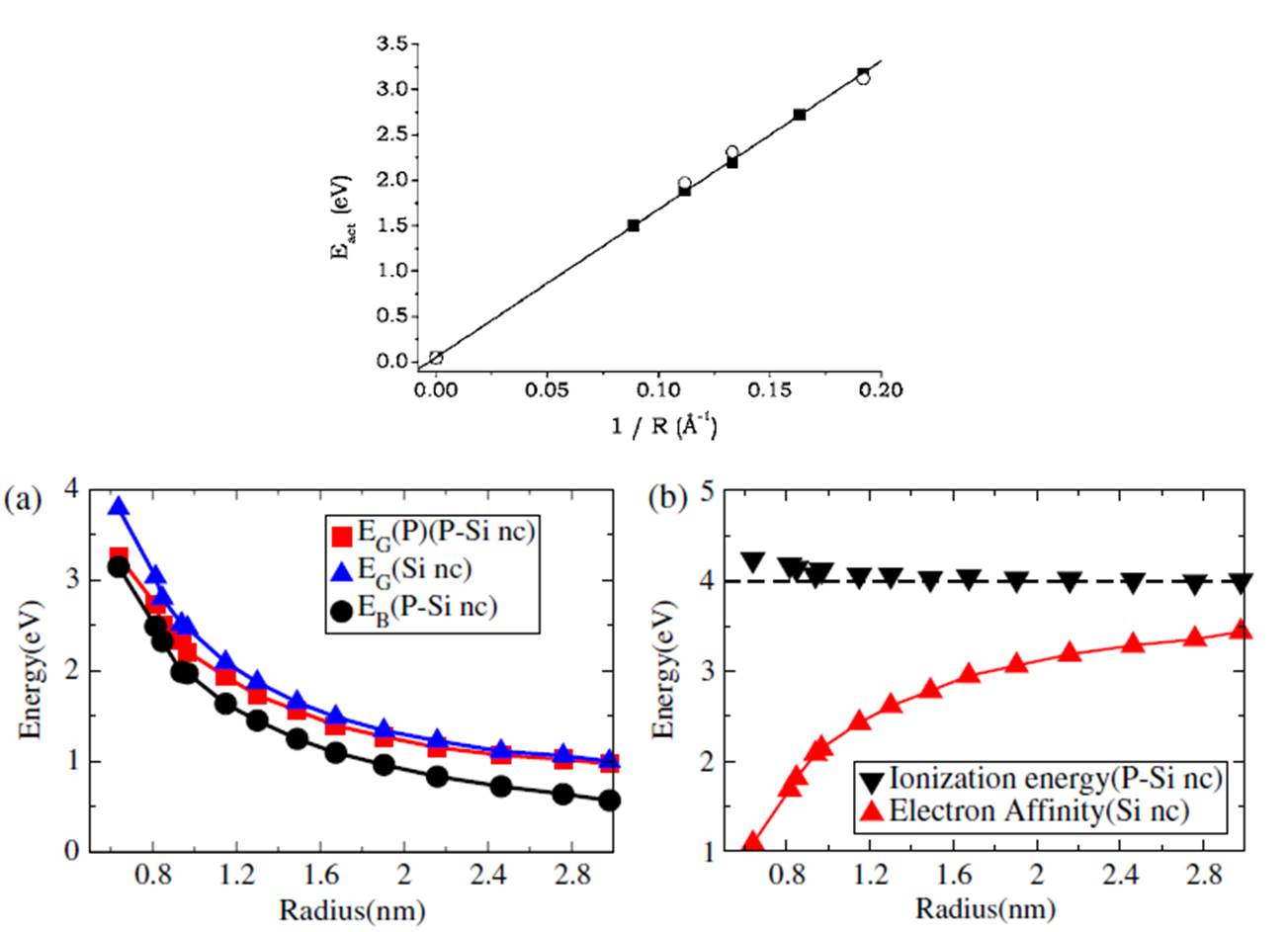}
\caption{(Color online). Top Panel: impurity activation energy, as a function of the inverse
Si-NC radius. Both B- (filled squares) and P-doped (open circles) Si-NC are considered. The zero inverse radius corresponds to bulk Si. The solid line is a linear fit of the data for the B-doped Si-NCs. The impurity is located at the Si-NC center. Reprinted with permission from Ref. \citep{CantelePRB2005}.
Bottom Panel: (a) the dependence on the Si-NC radius of the energy gap $E_G$ in Si-NCs (up triangle), of the activation energy or  binding energy $E_B$ (dots), of the energy difference between the defect level and the highest doubly occupied level $E{_G}(P)$ in P-doped Si-NCs (square). (b) Ionization energy of
P-doped Si-NCs (down triangle) and electron affinity of undoped Si-NCs (up triangle) plotted as a function of Si-NC radius. Reprinted with permission from \citep{ChelikowskyRPP2011}.}
\label{fig:figure8}
\end{center}
\end{figure}

Calculated $E_{act}$ for B- and P-doped Si-NCs, with dopants located at the nanocrystal center, as a function of the Si-NC size are shown in the top panel of Fig. \ref{fig:figure8}. The figure shows that $E_{act}$  scales almost linearly with the inverse of the Si-NC radius \citep{CantelePRB2005}, indicating that the main contribution to $E_{act}$ is mainly due to the almost unscreened Coulomb interaction.
This is a consequence of the reduction of the screening that is observable moving to the nanoscale \citep{OgutPRL1997} and that also leads to a strong localization  of the impurity states \citep{MelnikovPRL2004}. The calculated activation energies for P-doped Si-NCs \citep{CantelePRB2005} are in fair agreement
with those calculated in Ref. \citep{MelnikovPRL2004} and showed in the bottom panel of Fig. \ref{fig:figure8}. 

Here it is interesting to note that the behavior of the activation energy is mainly due to the decrease with size of the Si-NC electronic affinity, whereas the ionization energy depends weakly on Si-NC size. It is worthwhile to note that investigations by magnetic resonance of the P hyperfine splitting in doped Si-NCs \citep{PereiraPRB2009} showed that the donor localization is dominated  by a reduction in dielectric screening  for Si-NCs with radii above 6 nm, whereas for radii below 2 nm quantum confinement dominates. In the intermediate region  both dielectric and quantum confinement contribute to the donor wave-function localization. Finally a screening model, based on the Thomas-Fermi theory has highlighted the importance of surface polarization effects. The activation energies of the impurities calculated with this model are in close agreement with the ones obtained using the 
most computational demanding ab-initio  procedures \citep{NinnoEL2006,NinnoPRB2006,TraniPE2009}. 
All these evidences show that in Si-NCs the impurity activation energies are quite large with respect to their Si bulk  value (0.045 eV for both B and P), demonstrating that the impurity ionization in Si-NCs is strongly quenched with respect to the bulk, in agreement with the experiments \cite{PolisskiPB1999,LechnerJAP2008,StegnerPRL2008}.  The activation energy rapidly decreases by increasing the Si-NC size, as confirmed by the activation of P-dopants for large Si-NCs, observed by Kelvin force probe microscope (KFPM) \citep{XuJAPP2014}. 

The calculated dopant activation energy  depends not only on the Si-NC size, but also on the position of the dopant atom in the Si-NC. On going from the Si-NC center to the Si-NC surface, $E_{act}$ shows variations up to 0.2 eV  \citep{MelnikovPRL2004,CantelePRB2005,ChanNL2008,MavrosJPCC2011}. Moreover
by increasing doping concentration, $E_{act}$ strongly increases, as a consequence of the enhanced structural relaxation induced by the dopants in the Si-NC
\citep{PiJAP2014}. 

The knowledge of the impurity wave-function localization allows the calculation of the hyperfine splittings (HFS), which can detect the presence of dopants in electron paramagnetic resonance (EPR) experiments. Calculations of the HFS for P-doped Si-NCs with diameter up to 6 nm (impurity at the Si-NC center) \citep{ChanNL2008} show a good agreement with the experimental results \citep{FujiiPRL2002,SumidaJAP2007}. Moreover the variations of the HFS calculated values with respect to the dopant position in Si-NCs give another indication of the possibility to obtain Si-NCs with P impurity localized in the Si-NC core.

It is interesting to note that, in the calculations of the vibrational properties and Raman spectra of P-doped Si-NCs \citep{KhooPRB2014}, it has been found  that i) the Si-NC vibrational density of states is only slightly modified by the presence of P-dopant atoms, regardless of its location, ii) the Raman spectra of doped Si-NCs are enhanced with respect to the undoped ones and are strongly dependent on the P-dopant position. Raman spectroscopy may, thus, be used as a method for characterizing not only Si-NC diameter, but also dopant location within the Si-NCs.

The theoretical results for the optical properties reflect the outcomes relative to the energy level diagrams \citep{ZhouACS2015,ZhouJACS2003,RamosJPCM2007,MavrosJPCC2011,PiJAP2014,RamosPRB2008,PiPRL2013}.
Since group III and group V doped  Si-NCs have an odd number of electrons, usually spin polarized calculations are performed. The relative position of the defect energy level, calculated with respect to the HOMO and LUMO states of the undoped NC, strongly depend on the Si-NCs size. In particular, an increase of the QC, which means a decrease of the Si-NC size, leads to the formation of deeper impurity levels \cite{RamosJPCM2007,XuPRB2007}. Moreover, differently from the bulk, the extra hole, or electron, tends to remain localized near the impurity\cite{MelnikovPRL2004,IoriPRB2007}. The impurity level with the lower energy (spin-up or spin-down) is
occupied, while the level with higher energy is not occupied.
 Besides, the increase of the QC induced by size reduction leads to an enhancement of the differences in energy between the spin-up and spin-down impurity-related levels. 

\begin{figure}
\begin{center}
\includegraphics[width=12.cm,height=9.64cm]{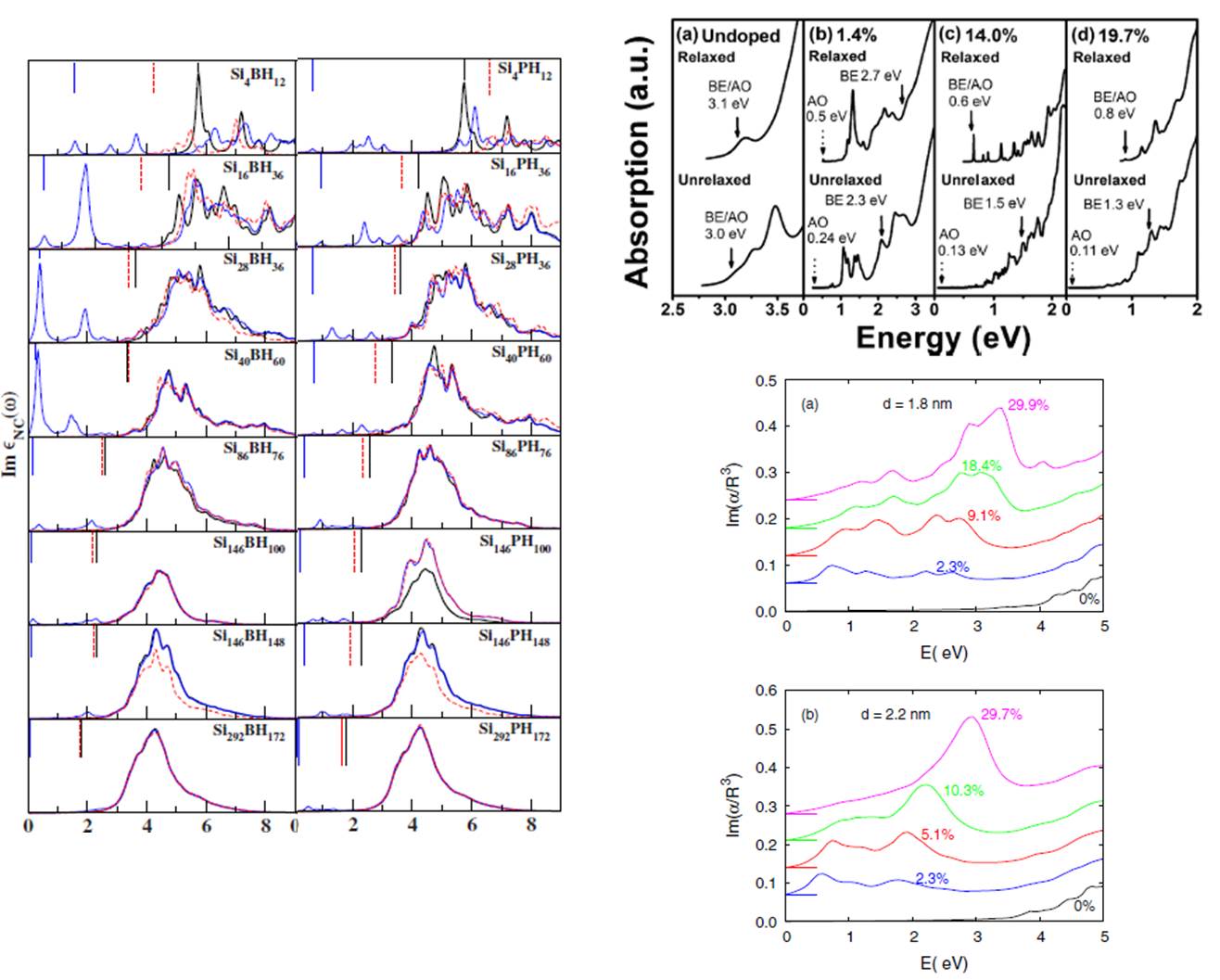}
\caption{(Color online). Left Panel: optical absorption spectra and energetic positions of the lowest optical transitions (vertical lines from the top) for the undoped Si-NCs (thick solid lines) and of Si-NCs doped with B and P impurities in their neutral (thin solid line) and ionized (dashed line) charged states. Reprinted with permission from Ref. \citep{RamosPRB2008}. Right Top Panel: optical absorption spectra for (a) an undoped Si$_{71}$H$_{84}$-NC  and Si-NCs doped with P at the concentration of (b) 1.4\%, (c) 14.0\% and (d) 19.7\% both before and after structural relaxation. Calculations are performed within TDDFT. The energy that corresponds to a band-edge (BE) transition is indicated by a solid arrow. Each absorption onset (AO) is indicated by a dashed arrow. Reprinted with permission form Ref.  \citep{PiJAP2014}. Right Bottom panel : absorption spectra of (a) 1.8 nm Si-NCs (Si$_{87}$) doped with 0, 2, 8, 16 and 26 P atoms. (b) Same for 2.0 nm Si-NCs (Si$_{175}$) doped with 0, 4, 9, 18 and 52 P atoms. The corresponding doping concentrations are also indicated. The increase of the P concentration results in a broad peak, that becomes progressively prominent as the extra electrons brought by the P atoms initiate to respond collectively to the external field. Reprinted with permission from \citep{PiPRL2013}.}
\label{fig:figure9}
\end{center}
\end{figure}

Figure \ref{fig:figure9} collects some theoretical results concerning the optical absorption properties. On the left we see that the lowest transition energy of the neutral B-impurity tends to zero, as the size of the doped Si-NCs increase.  Moreover, the oscillator strength of this transition tends to decrease for large doped Si-NCs. The values of the lowest transition energies of the Si-NCs doped with the ionized B-impurity approach the lowest transition energy of the undoped Si-NCs. For the largest B-doped Si-NC the absorption spectrum looks practically the same of the undoped Si-NC. This shows that changes in the optical spectra of B-doped Si-NCs should be caused by neutral impurities. 
PL spectra due to the transition from the conduction band to the impurity level have been observed for B-doped free-standing Si-NCs (see discussion above regarding the experimental outcomes). Concerning P-doped Si-NCs the impurity-related peaks in the
absorption are not so intense as in the case of B-doped Si-NCs. Moreover the values of the lowest transition energies of the Si-NCs doped with an ionized P atom are larger than the lowest transition energy of the Si-NCs doped with a neutral P impurity and resemble the ones of undoped Si-NCs.  For large Si-NCs the role of the P doping is predicted to be more important with respect to that of B doping since the B-related peaks practically vanish for the largest Si-NC, in contrast to the P-related peaks. 

Spin polarization plays clearly a role on the spectra of doped Si-NCs. Since the number of electrons in the two groups of spin states differs by one, the HOMO and LUMO of each group of states differ in energy. The minority-spin group of states has one electron less and its HOMO and LUMO are very close in energy, whereas the HOMO-LUMO energy difference of the majority-spin states corresponds roughly to the energy difference between the impurity level and the LUMO of the undoped system. The HOMO-LUMO transition of the minority-spin states presents a very high oscillator strength, and the transitions involving the impurity state and states below the HOMO give rise to intense peaks in the absorption. On the other hand, the spectra corresponding to the majority-spin states are very similar to the one of the undoped Si-NCs  \citep{RamosPRB2008}. 

Also the dopants concentration and the induced structural deformation play an important role regarding the optical properties of doped Si-NCs. By means of a TDDFT study it has been showed (see top right in Fig. \ref{fig:figure9}) that for structurally unrelaxed P-doped Si-NCs, sub-band absorption, related to electronic excitations from impurity, is induced by the presence of deep energy levels in the low energy region \citep{PiJAP2014}. By increasing dopant concentration both the absorption onset and the band-edge absorption red-shift. Experimentally infra-red (IR) absorption has been observed in P-doped Si-NCs togheter with a red-shift of the PL \citep{ImakitaOE2009}. After structural relaxation of the system \citep{PiJAP2014} the situation does not change very much for lower dopant concentration. Anyway, an increase in the dopant concentration induces increasing structural relaxation, that remove deep impurity states from the band gap region (i.e. absorption onset and band-edge absorption coincide). Moreover a slight blue-shift is calculated at larger concentration. Tight-binding calculations 
\citep{PiPRL2013} pointed out another interesting consequences of hyperdoping. By calculating the optical response of Si-NCs ideally doped with different and large concentrations of P atoms, a collective response of P-induced electrons was proven. This leads to localized surface plasmon resonance phenomena, ideal for the development of Si-based plasmonic devices, enabling silicon-compatible IR photonics at the nanoscale  \citep{ZhouACS2015,RoweNL2013}. 

\section{Matrix-embedded Nanocrystals}
\label{sec:matr}

Matrix-embedded Si-NCs offer advantages in terms of stability and low-cost manufacturability. Moreover they can be exploited to develop CMOS compatible devices and novel third-generation photovoltaic systems. Methods for the fabrication of matrix-embedded Si-NCs are the  plasma enhanced chemical vapor deposition (PECVD) \citep{Iaconabook2010}, the reactive magnetron sputtering \citep{Gourbilleaubook2010}, the ion implantation \citep{Ellimanbook2010} and the  formation of superlattices of alternating SiO$_x$ and SiO$_2$ layers (see Fig. \ref{fig:figure0}, for scheme of the  different techniques used)
\citep{GourbilleauJAP2009,ZachariasAPL2002,KhriachtchevJAP2002}. 
Unfortunately, the high resistivity of the embedding matrix reduces the possibility to employ such materials in applications that require a high carrier mobility. For this reason, the possibility to introduce dopants has been explored in order to improve transport and other fundamental characteristics \citep{OlivaNano2016,ArducaMSSP2017}. 

For matrix-embedded Si-NCs the doping of nanostructures involves  additional complications arising from the presence of both interfaces and materials with different affinities in hosting  the dopant species \citep{NorrisScience2008}. In practice, experimental samples of matrix-embedded Si-NCs present inhomogeneous impurity concentrations that depend on the fabrication technique. In other words, since dopants tend to diffuse during the fabrication process, one cannot simply limit the doping to a selected region of the material. Therefore, in order to connect some physical results to the presence of dopants, we have  to understand where the impurities are located, that is if they are in the nanostructures, at the interfaces or in the embedding matrix.
 To accomplish this, many experimental techniques have been employed, together with ab-initio calculations.\\
 \noindent
Fabrication of doped  Si-NCs follows two main procedures. In the first one, doping occurs during the Si-NCs synthesis, i.e. the dopants are incorporated without additional steps. In the second one, instead, doping occurs after the Si-NCs synthesis. 
\begin{figure}
\begin{center}
\includegraphics[width=12.0cm,height=6.68cm]{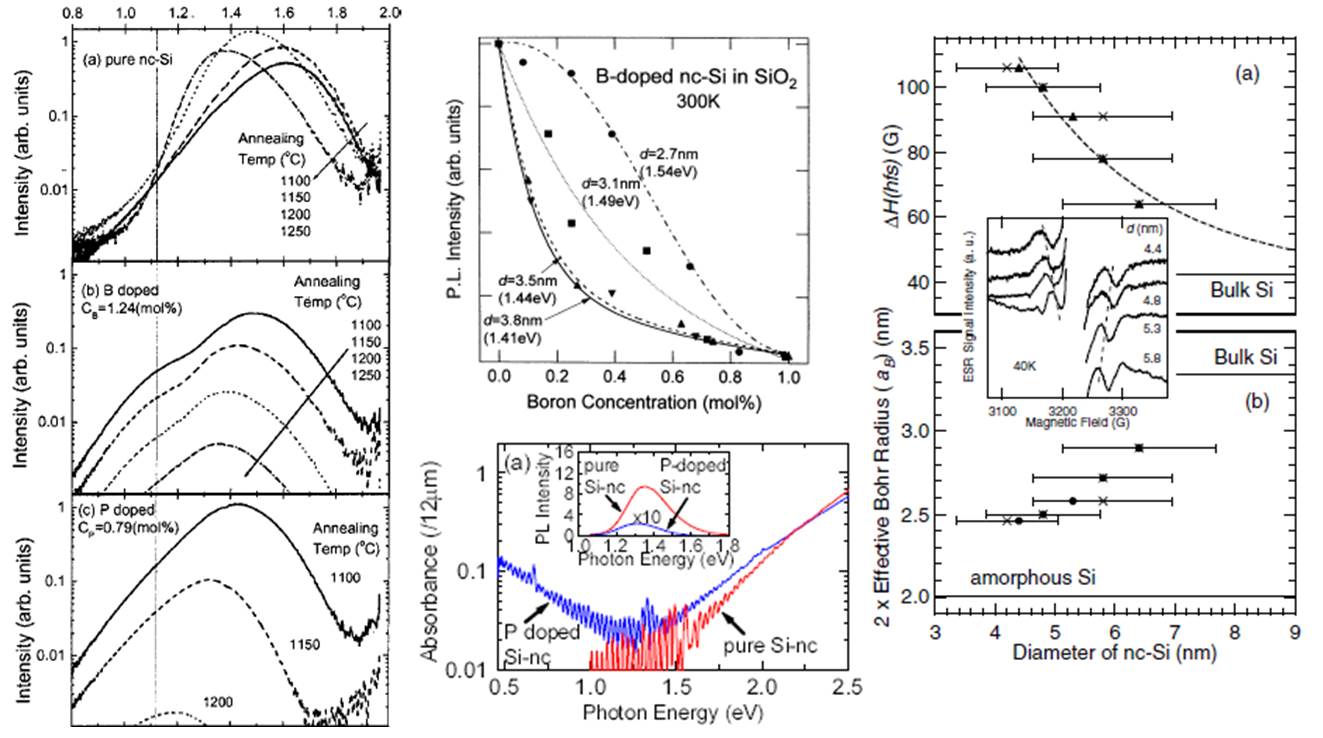}
\caption{(Color online) Left panel: annealing temperature dependence of the PL spectra of (a) undoped Si-NCs,(b) B-doped Si-NCs and (c) P-doped Si-NCs. All spectra are taken under the same conditions. The vertical line indicates the band gap energy of bulk Si crystals (1.12 eV). Reprinted with permission from Ref. \citep{FujiiAPL2004}. Top Center Panel: dependence of PL peak intensity as a function of B concentration for Si-NCs with diameter from 2.7 to 3.8 nm. The values in parentheses represent the PL peak energies. All the results are normalized to their maximum intensities for comparison purposes. The curves are drawn to guide the eye. Reprinted with permission from Ref. \citep{MimuraSSC1999}. Bottom Center Panel: absorption spectra of undoped and P-doped Si-NCs (impurity concentration 0.8 mol$\%$). The inset shows the PL spectra of the same samples. Reprinted with permission from Ref. \citep{ImakitaOE2009}. Left Panel: (a) P hyperfine splitting and (b) twice the value of the effective Bohr radius as a function of Si-NCs diameter. P concentration is within the range 0.2 and 0.8 mol$\%$. The inset is the HFS of ESR derivative spectra for the samples with different Si-NC sizes. Reprinted with permission from Ref. \citep{FujiiPRL2002}.}
\label{fig:figureexp3}
\end{center}
\end{figure}

The first report of B and P doping performed during Si-NCs synthesis was due to the group of Fujii \citep{KanzawaSSC1996}. Si, SiO$_2$, P$_2$O$_5$ or B$_2$O$_3$ were simultaneosuly sputter-deposited and then annealed in N$_2$ gas atmosphere. During the annealing the B- or P-doped Si-NCs were grown in BSG or PSG, i e. B- or P-doped silica, glass matrices 
\citep{ImakitaOE2009,FujiiAPL2005,AlmeidaPRB2016,NomotoJPCC2016,NomotoMRS2016,FujiiPRL2002,SumidaJAP2007,FujiiAPL2004,MimuraSSC1999,FujiiAPL1999,MimuraJLum1999,MimuraPRB2000,FujioAPL2008,FujiiJAP1998}. The doped Si-NCs were investigated by HRTEM \citep{MimuraSSC1999,KanzawaSSC1996,MimuraPRB2000,FujiiJAP1998}, Raman\citep{KanzawaSSC1996}, ESR \citep{AlmeidaPRB2016,FujiiPRL2002,SumidaJAP2007,FujiiJAP2000} absorption and PL spectroscopies \citep{ImakitaOE2009,FujiiAPL2005,FujiiPRL2002,SumidaJAP2007,FujiiAPL2004,MimuraSSC1999,MimuraPRB2000,FujiiJAP1998,FujiiJAP2000}, ATP and proximity histograms, that enable the visualization of the Si-NCs structure as well as of the dopant distribution with subnanometer resolution \citep{NomotoJPCC2016}. 

The obtained results pointed out that the presence of impurities during the synthesis affects the growth kinetics of the Si-NCs, resulting in a wide size distribution that depends on the annealing temperature and on the nature and concentration of dopants. Doped Si-NCs with a diameter ranging from 2.7 to 3.8 nm, slightly larger than the corresponding undoped Si-NCs grown using the same annealing temperature, were prepared \citep{ArducaMSSP2017}. The doped Si-NCs showed good crystallinity and were well separated from each other in the matrix. The PL spectra showed a quenching in the intensity (top center panel of Fig \ref{fig:figureexp3}). For B doping, the quenching was found to increase with the NC size. This behavior was explained by assuming that B impurities were localized  in substitutional sites in the Si-NCs and that they were responsible to promote non-radiative recombination mechanisms (Auger recombination) of  trions formed by the photogenerated e-h pairs and the B-induced hole.
Concerning the PL activity in P-doped systems, two opposite trends were recorded. Initially an increase in the PL intensity was detected, then, on increasing the P concentration, a decrease of the PL intensity was recorded. Also in this case this PL quenching  was explained in term of Auger recombination, whereas the initial PL intensity increase was assigned to P passivation of the dangling bonds at the interface between the Si-NCs and the matrix. 

In both cases, the PL spectra became broader, indicating the development of low energy features with respect to the undoped case (left panel of Fig. \ref{fig:figureexp3}). Optical absorption in the infrared region, was observed for  P-doped Si-NCs. In this region of frequencies, absorption increases when the P concentration increase and seems to be not affected by the presence of the PSG matrix. This effect was interpreted as the consequence of intra-conduction band transitions (bottom center panel of Fig  \ref{fig:figureexp3}) induced by the presence of the impurities.
 Quantum confinement of P donors in Si-NCs was evidenced through the clear dependence of the hyperfine splitting on the Si-NC size (left panel of Fig \ref{fig:figureexp3}). Regarding the impurity location, APT and proximity histogram \citep{NomotoJPCC2016} revealed incorporation of B and P into the Si-NCs with clear differences. The B concentration sligthly decreased on going from the matrix to the Si-NC interfaces and from the Si-NC interfaces to the NC core. The P concentration started to increase about 1 nm far from the Si-NC interface, reaching a maximum in Si-NC core (see bottom panel of Fig \ref{fig:figureexp2}).   

B- and P-doped Si-NCs formed as multilayers in a SiO$_2$ matrix have been fabricated, aiming at the development of silicon quantum dot based solar cells 
\citep{ConibeerPPRA2011,HaoNanoT2009,NomotoMRS2016,HaoNanoT2008,PerezAPL2009,HaoTSF2009,
HaoSESC2009a,HaoSESC2009b,XieAPL2013,VeettilAPL2014,ZhangJAP2015}. B-doped Si-NCs were formed by magnetron co-sputtering of Si, SiO$_2$ and B$_2$O$_3$ targets, while for P-doped Si-NCs, Si, SiO$_2$ and P$_2$O$_5$ targets were used. Doped Si-NCs were fabricated by PECVD using the superlattice approach, that permits to well control the size and the distribution of the Si-NCs \citep{PeregoNANO2010,PeregoSIA2013,NomotoPSS2017,GutschAPL2012,GnaserJAP2014,
GutschAPL2015,KonigSR2015,LuSR2016,LuOME2016,QianNRL2016,HillerSR2017,
MastromatteoSIA2014,PeregoNANO2015,MastromatteoJMCC2016,YuSM2015}. 
The dopants location and the doping effects were studied through SIMS 
\citep{HaoSESC2009a,HaoSESC2009b,GutschAPL2012} and time-of-flight-SIMS (ToF-SIMS) \citep{PeregoNANO2010,MastromatteoSIA2014,MastromatteoJMCC2016}, XPS \citep{PeregoNANO2010,PeregoSIA2013,HaoNanoT2008,QianNRL2016,PeregoNANO2015}, FTIR   
\citep{HaoNanoT2009,HaoNanoT2008,HaoTSF2009}, TEM and HRTEM \citep{ConibeerPPRA2011,PeregoNANO2010,HaoNanoT2008,HaoTSF2009,
XieAPL2013,LuSR2016,QianNRL2016,MastromatteoSIA2014,PeregoNANO2015,YuSM2015}, ESR \citep{VeettilAPL2014,LuSR2016}, glancing incidence x-ray diffraction (GIXRD)  \citep{ConibeerPPRA2011,HaoNanoT2008,ZhangJAP2015}, absorption and PL spectra \citep{ConibeerPPRA2011,PeregoNANO2010,HaoNanoT2008,HaoTSF2009,VeettilAPL2014,ZhangJAP2015,
GutschAPL2012,LuSR2016,LuOME2016,HillerSR2017,YuSM2015}, XPS \citep{HaoTSF2009,XieAPL2013,LuSR2016,QianNRL2016}, XANES \citep{KonigSR2015}, Raman \citep{XieAPL2013,LuSR2016,YuSM2015}, Rutherford backscattering spectrometry (RBS) \citep{MastromatteoSIA2014} and APT and proximity histogram analysis \citep{NomotoMRS2016,GnaserJAP2014,KonigSR2015,HillerSR2017}.

\begin{figure}
\begin{center}
\includegraphics[width=10.0cm,height=11.73cm]{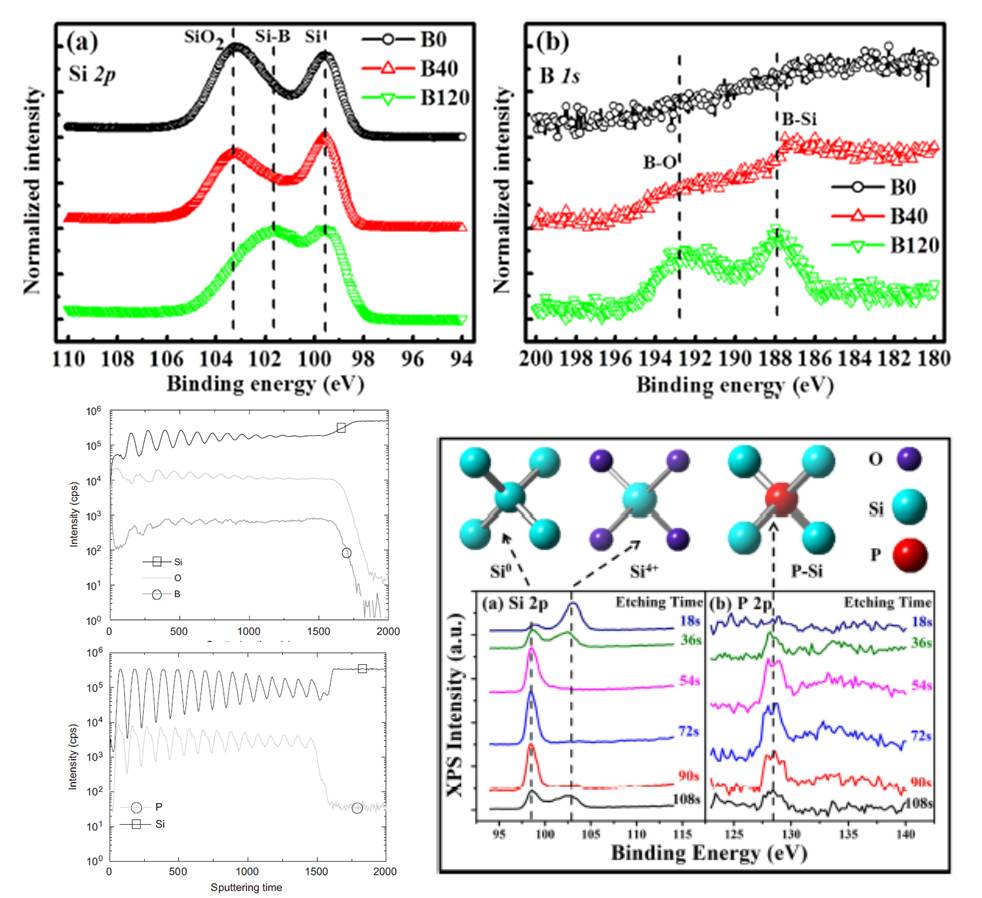}
\caption{(Color online) Top panel: (a) Si 2p core level XPS spectra of undoped silicon rich silicon oxide (SRSO) films (B0, top curve), and differently B-doped SRSO films (B40, middle curve) and B120 (bottom curve). (b) B 1 s core level XPS spectra of B0 (top curve), B40 (middle curve) and B120 (bottom curve). Reprinted with permission from Ref. \citep{XieAPL2013}. Bottom left panel: SIMS data of B-doped Si-NCs and SIMS data of P-doped Si-NCs, reprinted with permission from \citep{HaoSESC2009a} and \citep{HaoSESC2009b}. Bottom right panel: depth-profile XPS spectra. The nominal P concentration is 2$\%$. (a) Si 2p and (b) P 2 peak peaks for the P-doped Si-NCs/SiO$_2$ multilayers after annealing are detected. Insets are the schematic diagrams of Si-Si, Si-O and Si-P bonds. Reprinted with permission from Ref. \citep{LuSR2016}.}
\label{fig:figureexp3a}
\end{center}
\end{figure}

\begin{figure}
\begin{center}
\includegraphics[width=12.0cm,height=8.19cm]{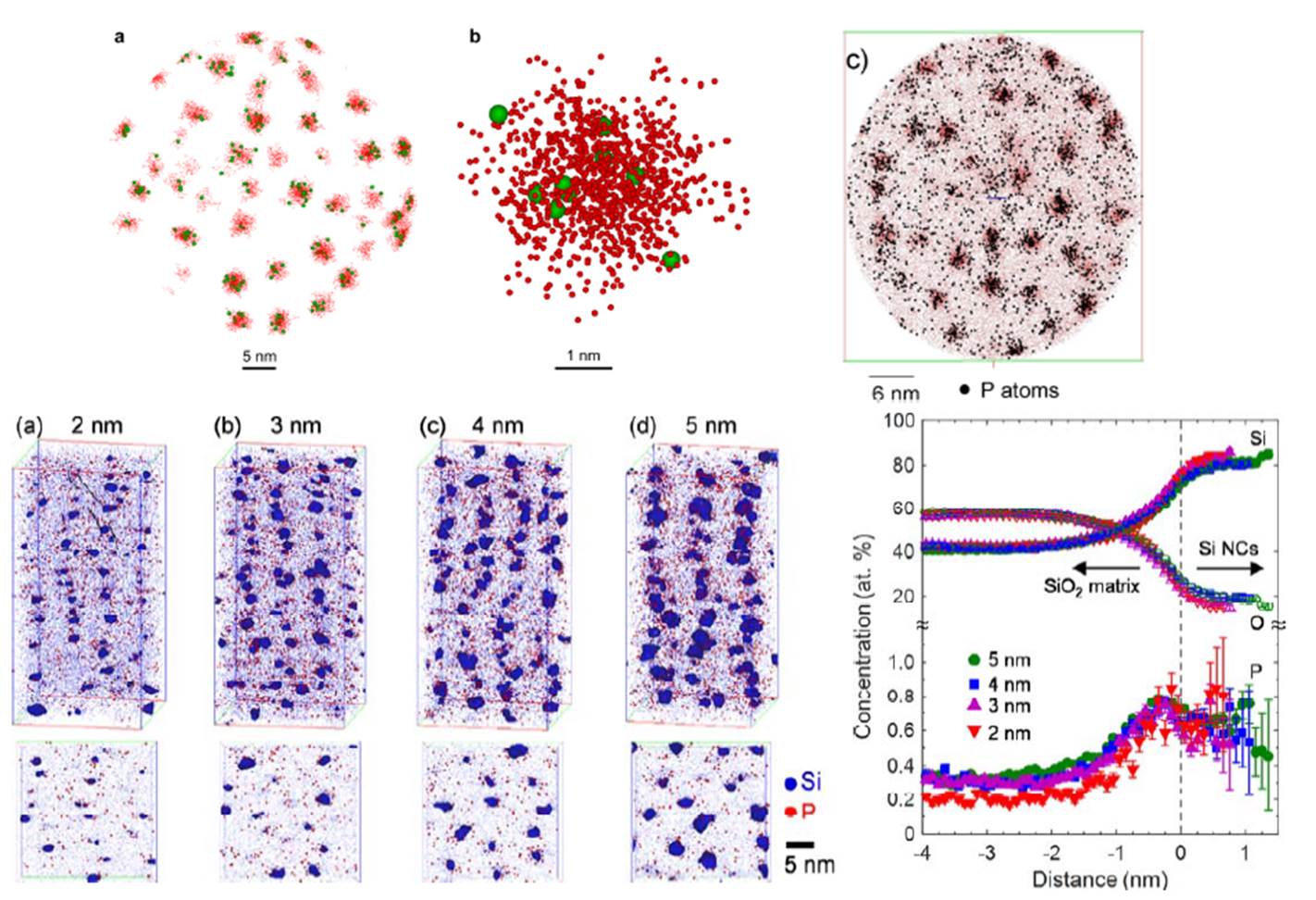}
\caption{(Color online) Top Panel: (a) a 3D representation of individual silicon clusters identified in the cluster analysis in a slice (8 nm thick) perpendicular to the z-direction from one of the silicon rich oxynitride (SRON) layers. Individual clusters show Si atoms (red dots) and P atoms (green circles). (b) An individual Si-NC, containing 970 Si atoms (red circles) and 9 P atoms (green circles). Reprinted with permission from Ref. \citep{GnaserJAP2014}. (c) Plane view of thin slice extracted from the atom probe tomography of a Si+P implanted sample: $^{31}$P atoms (black dots) are shown in addition to $^{28}$Si atoms, evidencing the presence of a high quantity of P inside the Si clusters. Reprinted with permission from \citep{KhelifiAPL2013}. Bottom Panel Left: APT reconstructions of P-doped Si-NCs with SRON:P thicknesses of (a) 2 nm, (b) 3 nm, (c) 4 nm, (d) 5 nm. In the upper row the superlattice stacking order is from left to right and box sizes are 25x25x60 nm$^3$. In the lower row, projections through a tomographic slice of one superlattice layer of SRON:P (25x25x6 nm$^3$) are shown. Bottom Panel Right: proximity histogram analysis of the P-doped Si-NCs embedded in SiO$_2$. Samples with SRON:P tickness of 5 nm are shown as green (circles), 4 nm as blue (squares), 3 nm as purple (triangles) and 2 nm as red (inverted triangles). Reprinted with permission from Ref. \citep{NomotoPSS2017}.}
\label{fig:figureexp5}
\end{center}
\end{figure}

\begin{figure}
\begin{center}
\includegraphics[width=12.0cm,height=11.20cm]{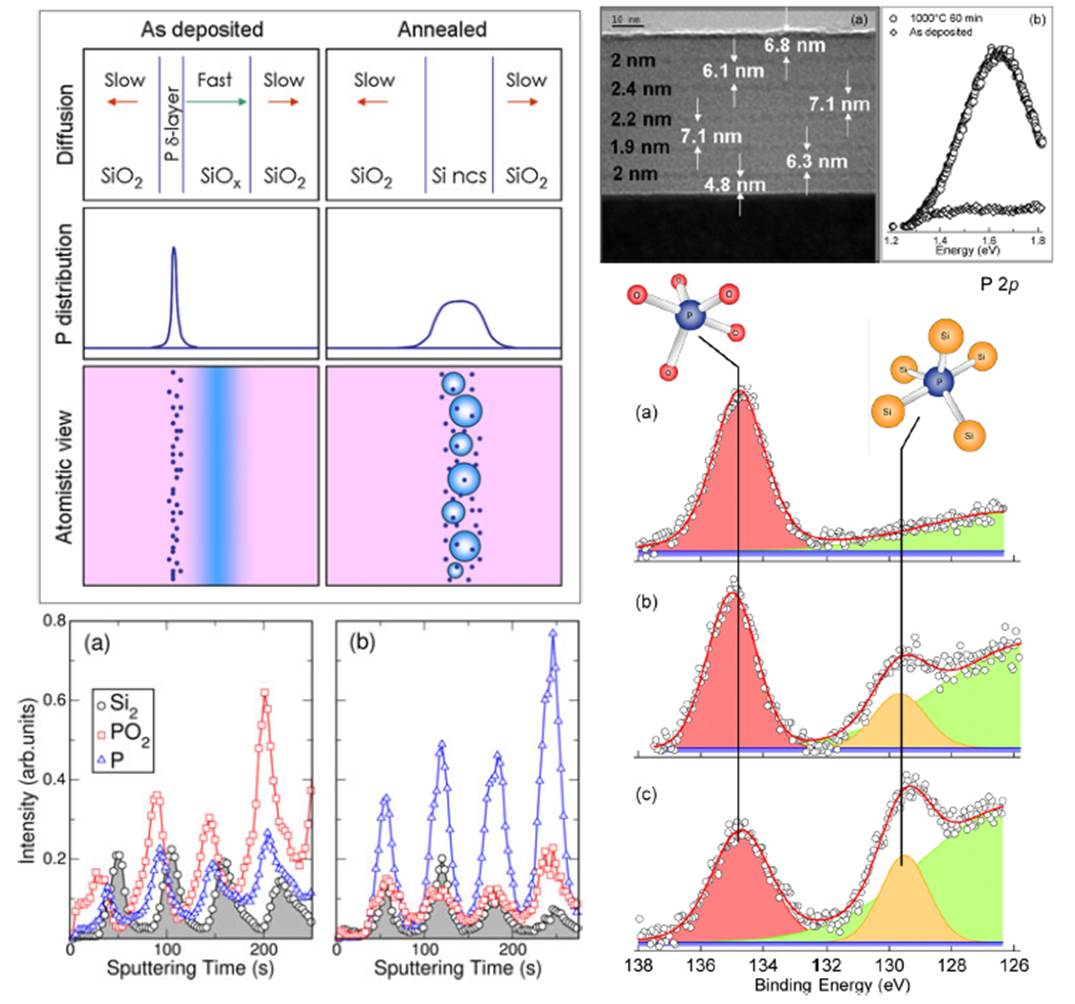}
\caption{(Color online) Top Left: schematic description of the P mechanisms governing redistribution of P atoms (dots) in Si/SiO$_2$ multilayer structures during high-temperature thermal treatments. Bottom Left: ToF-SIMS depth profile of the Si/SiO$_2$ multilayer structure, with interposed P-$\delta$-layer of different thickness, before (a) and after (b) the thermal treatment in N$_2$ atmosphere. Reprinted with permission from Ref. \citep{PeregoSIA2013}. Top Right: (a) TEM image of the SiO/SiO$_2$ multilayer structure after annealing. The dark contrast corresponds to silicon rich regions. (b) PL spectra of the as-deposited and annealed sample. Reprinted with permission from Ref. \citep{PeregoNANO2010}. Bottom Right Panel: high resolution XPS spectra of the P 2 p region for the as deposited and two annealed samples. Si-NCs with average diameter equal or smaller than 2 nm are arranged in a superlattice. Reprinted with permission from Ref. \citep{PeregoNANO2010}.}
\label{fig:figureexp4}
\end{center}
\end{figure}

The behavior of the PL  spectra was found to be similar to those of the doped Si-NCs embedded in BSG and PSG.  SIMS data suggested that the dopants were located in the silicon layer and not in  the silica layer \citep{HaoNanoT2009,HaoNanoT2008} (see Fig. \ref{fig:figureexp3a}, bottom left panel), while  XPS data pointed out that B is either inside or at the surface of the Si-NCs, with a concentration equilibrium distribution of B between the Si and the SiO$_2$  \citep{XieAPL2013} (see Fig. \ref{fig:figureexp3a}, top panel). In the case of P, the XPS spectra suggested that most of the P atoms were located in the Si layer and at the Si/SiO$_2$ interface \citep{LuSR2016} (see Fig. \ref{fig:figureexp3a}, bottom right panel). Even after high temperature annealing, P atoms did not tend to move into the SiO$_2$ layer and were better incorporated in the Si-NCs \citep{PeregoSIA2013,MastromatteoSIA2014,PeregoNANO2015}.  
APT and proxigram analysis were used to determine the dopants localization.  Gnaser et al. \citep{GnaserJAP2014} founded that the concentration of P was enhanced in the region of Si-NCs with a maximum at the Si/SiO$_2$ interface; their results seemed to indicate the occurrence of self-purifications effects.  Nevertheless the incorporation of P atoms in very small Si-NCs with diameter as small as 2 nm was demonstrated by the same group \citep{NomotoPSS2017} (see Fig. \ref{fig:figureexp5}). 
This result was confirmed by Perego et al. \citep{PeregoNANO2010,PeregoSIA2013} that,  using a combination of TEM, ToF-SIMS and XPS, showed that P atoms are definitevely incorporated within Si-NCs of diameter of the order of 2 nm. These results have been interpreted as a consequence of the favored P diffusion towards the Si-rich region (see Fig. \ref{fig:figureexp4}).  

Indeed it has been demonstrated that ion beam synthesis, which is a well known approach to grow undoped Si-NCs embedded in SiO$_2$, is also an efficient technique to grow doped Si-NCs with diameters of few nanometers, if the dopants are co-implanted with Si  \citep{KhelifiAPL2013,FregnauxJAP2014,FregnauxPSSC2015}. Combining the results obtained by  APT, Raman, PL and PLE analysis, the authors concluded that P impurities tend to be efficiently incorporated in the Si-NC core whereas B atoms are mainly localized at the  Si-NC/SiO$_2$ interface (see Fig. \ref{fig:figureexp5}, top left panel) \citep{KhelifiAPL2013,FregnauxJAP2014,FregnauxPSSC2015}.

The electrical activity of matrix-embedded doped Si-NCs, monitored by current voltage measurements, demonstrated that majority carriers, originated from substitutional P-donors, must be generated to overcome significant P-ionization energies between 110 and 260 meV. In the absence of electrical fields at room temperature, however, no significant free carrier densities were detected \citep{HillerSR2017}. 

As a consequence, it has been concluded that only a small fraction of P-atoms are in substitutional sites within the Si-NCs, while the majority of them, reside in interstitial sites. These atoms are considered to be at the origin of the PL quenching that was previoulsy assigned to the occurance of Auger recombination mechanisms \citep{GutschAPL2015,KonigSR2015,HillerSR2017}. Interestingly Almeida et al. \citep{AlmeidaPRB2016} have studied both free and SiO$_2$ embedded P-doped Si-NCs using EPR spectroscopy. Their results showed that P dopant atoms are prevalently incorporated at substitutional sites and act as donors. Moreover the donor electron density decreased by several order of magnitude when  matrix-embedded Si-NCs were etched by HF to remove the SiO$_2$ matrix and were subsequently exposed to air. Once more, these results pointed out that doping efficiency depends not only on the dopant location but also on the Si-NCs environment and on the doping strategies. A marked difference in effective mobility for B-doped Si-NCs has been evidentiated in sample obtained by diffusion or by in-situ doping methods, respectively \cite{ZhangJAP2015}.     

Doping after synthesis is usually made by ion implantation in Si-NCs formed in SiO$_2$ matrices 
\citep{KachurinSEM2003,TchebotarevaJL2005,MurakamiJAP2009,EhrhardtJAP2016} or delivering a controlled amount of dopant atoms from a spatially separated diffusion source \citep{MastromatteoSIA2014,PeregoNANO2015,MastromatteoJMCC2016} (see Fig. \ref{fig:figureexp5}). The first strategy suffers of implantation damages. The second, instead, make possible to decouple equilibrium properties from kinetics effects, thus providing, for the first time, results that can be compared directly with the theoretical predictions. These experiments show that high levels of P impurities can be introduced in the Si-NC core, trapped with a binding energy of 0.9 eV.

\subsection{Ab-initio calculations. Formation energies and electronic properties}
\label{sub:formen1}
Contrary to the case of doping in H-terminated and OH-terminated Si-NCs, there are very few theoretical investigations regarding the study of doped Si-NCs embedded in a host matrix. The reason is related to the complicated models, characterized by a large number of atoms, that are needed to simulate these systems.

\begin{figure}
\begin{center}
\includegraphics[width=12.cm,height=11.22cm]{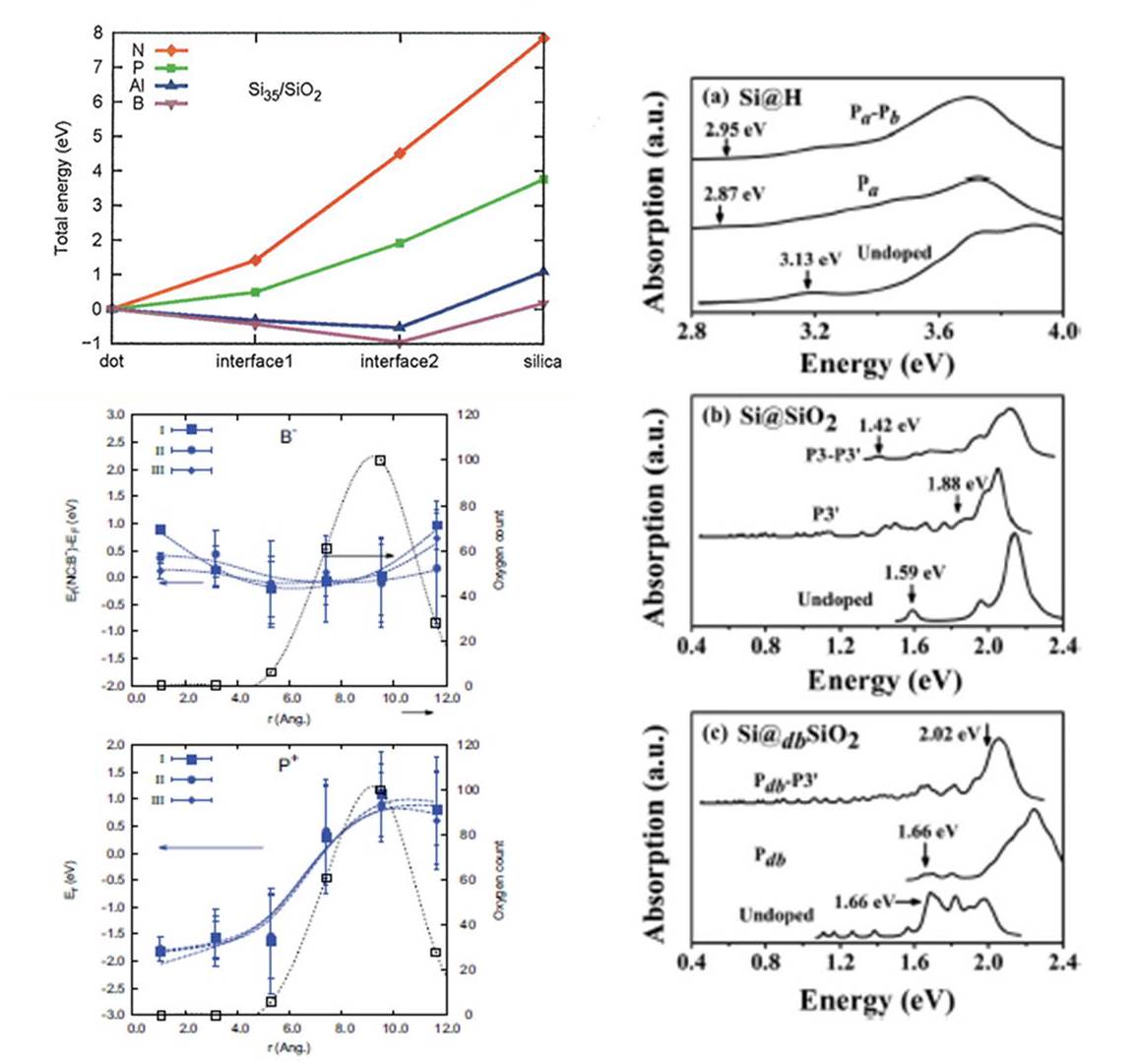}
\caption{(Color online). Top Panel Left: total energy of the Si$_{35}$-NC embedded in the SiO$_2$ matrix and doped with N, P, Al, or B, substitution of an atom in the Si-NC center (core), at the interface bonded to one or two O atoms (interface1, interface 2), or in the SiO$_2$ matrix far from the Si-NC (silica). For a better comprehension each line has been shifted so that the structure with the dopant in the Si-NC center has zero energy. Reprinted with permission from Ref. \citep{GuerraJACS2014}. Bottom Panel Left: relative energy of an SiO$_2$-covered nanoparticle doped with B (top) or P (bottom) as a function of distance from the centre. The filled symbols correspond to average formation energies for distance between $(r - 2$ ~\AA) and $(r + 2$ ~\AA), and the errorbars represent the standard deviation of the samples. Open squares represent the distribution of O (coincident for I, II and III). The relative energy is measured relative to the average for the samples. Reprinted with permission from Ref. \citep{CarvalhoPSSB2013}. Right Panel: optical absorption spectra of undoped, singularly, and doubly P-doped systems. (a) Si$_{71}$H$_{84}$, (b) Si$_{123}$O$_{96}$H$_{100}$ totally passivated, (c) Si$_{123}$O$_{96}$H$_{100}$ with a dangling bond. The excitation energy that corresponds to the transition from the top of the valence band to the bottom of the conduction band is indicated by the arrow. Reprinted with permission from Ref. \citep{NiPRB2014}}
\label{fig:figure10}
\end{center}
\end{figure}

Carvalho et al. \citep{CarvalhoPSSA2012,CarvalhoPSSB2013} modeled the embedded Si-NC considering an approximate spherical Si crystalline core of 1.5 nm  diameter, surrounded by a spherical shell of $\beta$-cristobalite SiO$_2$ with about 2 nm of outer diameter.  This Si-NC contained 161 Si atoms and 196 O atoms. To generate a realistic amorphous SiO$_2$ layer and a realistic Si-SiO$_2$ interface, the Si-NC was submitted to an anneal at 1800 K using molecular dynamics simulations. Three different annealing times were used, that is t = 2.0, 2.3 and 2.6 ps (the relative configurations are labelled with  I, II and III).
The structures obtained from the simulations were subsequently fully relaxed at 0 K, and all the dangling bonds were passivated by H atoms.
The new structures were then fully relaxed. The obtained systems were used to calculate electronic properties and dopants formation energies. The adopted procedure led to three different structures: the Si$_{161}$O$_{196}$H$_{59}$-NC for the annealing I and II and the  Si$_{161}$O$_{196}$H$_{57}$-NC for the annealing III.
As dopants either B (in their negative charge state) or P (in their positive charge state) were considered.  The preferred locations for subsitutional impurities were investigated calculating and comparing the energy of doped Si-NCs obtained by replacing B or P for each of the Si atoms in the three situations. The results showed that the doping behavior for Si-NCs covered by a SiO$_{2}$ shell differs from that of the corresponding H-terminated Si-NCs and that the behavior of P is different from that of B. Whereas the formation energy of B is nearly independent on the impurity location, with average formation energies only slightly lower at the interface, in the case of P doping the dopant has a strong preference to stay in the Si core and not in the SiO${_2}$ shell (see Fig. \ref{fig:figure10} bottom left panel). These outcomes were correlated to the fact that B atoms prefer to stay at interface locations where they form only three short Si-O bonds, while P atoms prefer to form a fourfold coordination with Si atoms. 

A different encapsulated Si-NC was obtained by Ni et al. \citep{NiPRB2014,NiPRB2017}. Starting from the H-terminated  Si$_{71}$H$_{84}$, they replaced H atoms with oxygens, which were then outwardly bonded by adding Si atoms. Finally, they passivated the system with H atoms to restore a fourfold coordination for all the Si atoms. The final system, that is the Si$_{123}$O$_{96}$H$_{100}$, represented a Si-NC covered by a  0.25-nm-thick SiO${_2}$ shell. Effects induced by the presence of  dangling bonds at the Si-SiO${_2}$ interface were also investigated.
The obtained results for the formation energies were in agreement with those of Carvalho et al. \citep{CarvalhoPSSA2012,CarvalhoPSSB2013} for the fully passivated Si-SiO$_2$ systems: B showed a small preference for the subinterface region, whereas P clearly preferred to stay in the Si-NC core. In contrast, when dangling bonds are induced in the system, B preferred to stay at the Si/SiO${_2}$ interface. It is important to note that the impurity formation energies are reduced on going from the H-saturated Si-NCs to the embedded ones. Moreover, electronic and optical properties  still revealed the presence of deep levels in the band gap and light emission at energies smaller than the energy gap of the undoped system
 \citep{NiPRB2014} (see Fig. \ref{fig:figure10} left panel).

Realistic calculations for Si-NCs  embedded in a solid matrix were performed by Guerra et al. \citep{GuerraJACS2014,NuriaNS2015}. SiO$_2$-embedded Si-NCs were generated starting from a 3x3x3 $\beta$-cristobalite-SiO$_2$ matrix (for a total of 648 atoms), by removing all the O atoms inside a sphere whose radius determines the Si-NC size. The so-obtained Si-NC, embedded in the SiO$_2$, presented perfectly coordinated atoms and the same Si-Si distance of $\beta$-cristobalite, corresponding to
about 3.2 \AA. After the relaxation, the Si/SiO$_2$ interface formed strongly stressed bonds, while far from the interface the bulk atomic densities were recovered \citep{GuerraPRB2009}. Dopant impurities were introduced in a substitutional Si-site located in the Si-NC center, at the Si/SiO$_2$ interface, or in the SiO$_2$ matrix far from the Si-NC. Since the Si atoms at tn interface can be bonded to one, two or three
O atoms (i.e., Si1+, Si2+, Si3+), for each structure all the possible oxidation types for the interfacial doping were considered. Each structure was fully relaxed after doping in order to include effects induced by structural rearrangements around the dopant atom.

In Fig. \ref{fig:figure10} (top left panel) the calculated total energies for a Si$_{35}$-NC (diameter of about 1 nm) embedded in SiO$_2$, single-doped with a group-V (N or P) or a group-III (Al or B) atom placed at
different substitutional sites, are shown. In the figure the minimum of each curve identifies the energetically favored site of the corresponding dopant. The data clearly indicate that for n-type doping (P and N) the impurity tends to settle in the Si-NC core, even for very small Si-NCs, confirming once more that dopant formation energies strongly depend on different interfaces. Instead, for p-type doping (Al and B) the interfacial sites are favored. Furthermore, it is clear that in all the considered cases the doping of the SiO$_2$ region is unlikely to occur. In particular, P doping requires a very high formation energy in order to move the dopant atom from the Si-NC core to the silica; this is in agreement with the observation that for Si-NCs embedded in a SiO$_2$ matrix, the matrix provides a strong barrier to P diffusion, inducing P segregation in the Si rich region \citep{PeregoSIA2013,PeregoNANO2015}. By contrast, still consistently with experiments \citep{XieAPL2013}, the diffusion of B toward SiO$_2$ is significantly easier. Once more the structural relaxation around the impurity plays a significative role. P atoms maintain a sort of tetrahedral coordination in all the cases, with all the bonds showing approximately constant characteristic lengths with Si and/or O. Instead, B atoms tend to form, when placed at the interface with two O atoms, a strong antibonding with one of the neighbors, causing repulsion to a large distance \citep{GuerraJACS2014}.

Several experimental results based on atom probe tomography and proximity histograms support these tendencies. APT demonstrated that n-type dopants (P and As) are efficiently introduced in the Si-NC core, whereas p-type dopants (B) are located at the Si-NC/SiO$_2$ interface 
\citep{KhelifiAPL2013,FregnauxJAP2014}. 
By proximity analysis Nomoto et al. \citep{NomotoJPCC2016} found significant differences in the B and P distribution profiles. For P-doped Si-NCs, the P concentration increases into the Si-NC region, whereas for B-doped Si-NCs the B concentration decreases moving from the 
Si-NC interface toward the Si-NC core. It is worthwhile to note that a similar picture has been observed also for doped Si nanowires (Si-NWs). In a joint experimental-theoretical effort it has been shown that, whereas the impurity B atoms are more stable at the Si-NW/SiO$_2$ interface, in a split interstitial configuration at a Si site, P impurities preferentially pile up in the cystalline Si region \citep{FukataJPCC2013}.

\begin{figure}
\begin{center}
\includegraphics[width=13.0cm,height=9.22cm]{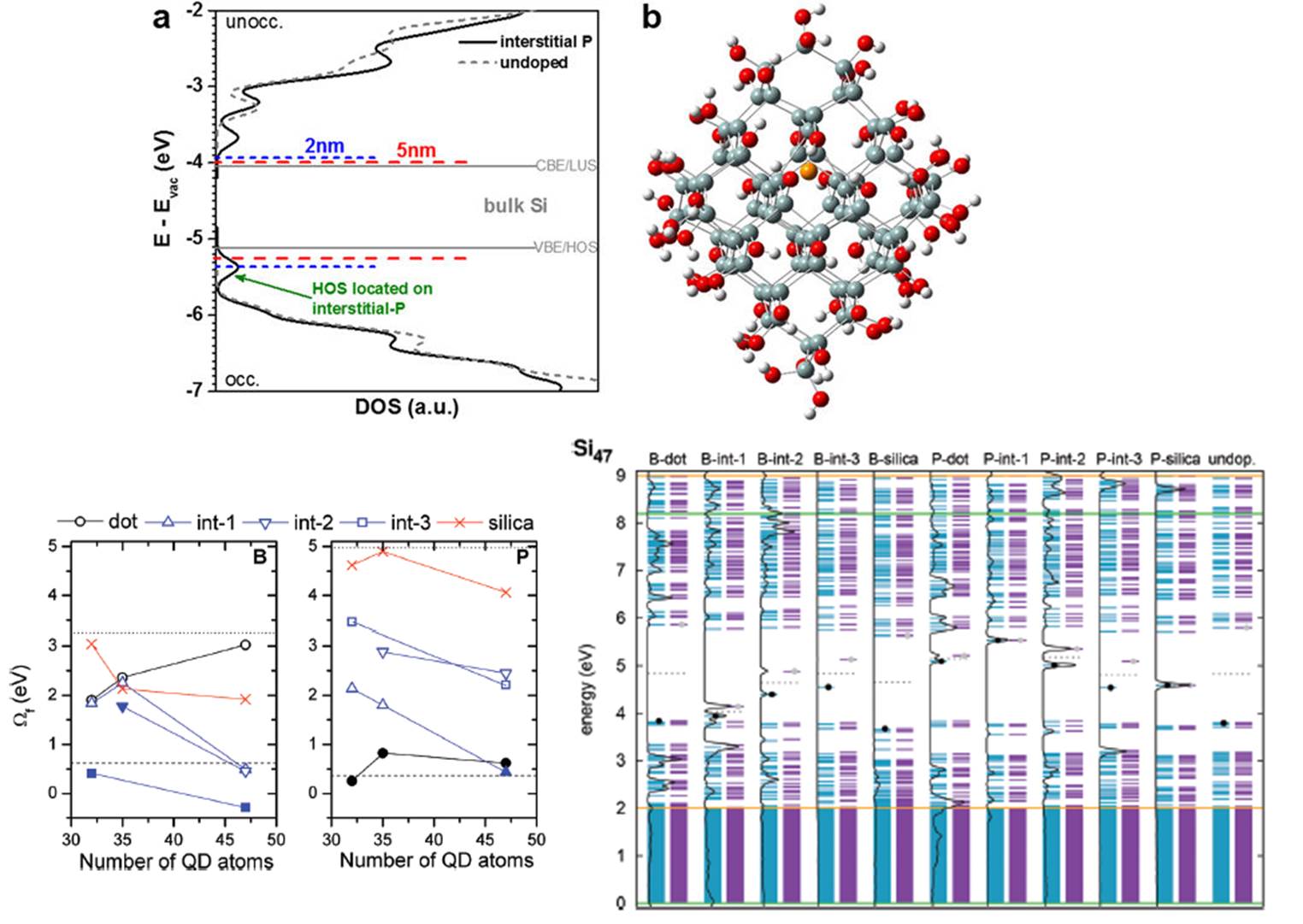}
\caption{(Color online) Top Panel: results of Density Functional Theory simulations. (a) Density of states (DOS) of pristine Si$_{84}$OH$_{64}$-NC (grey dotted curves, D = 1.5 nm) together with the DOS of a Si$_{44}$OH$_{64}$-NC containing interstitial P atom (black solid curves). The latter is shown in (b) with an interstitial P-atom in the Si-NC center (Si atoms: grey, O: atoms: red, H atoms: white, P atom: orange). For bulk Si (grey solid lines) the high occupied state (HOS) of interstitial P is well below the VB-edge and hence it is not a recombination centre. Quantum confinement induced widening of the fundamenal gap results in HOS energy of Si-NCs approaching the P-induced defect level. As a consequence small Si-NCs (blue dotted line) are more strongly subject to PL quenching than larger Si-NCs (red dashed line). Reprinted with permission from Ref. \citep{HillerSR2017}. Bottom: spin-up and spin-down eigenvalue spectra of doped Si$_{47}$-NC embedded in a SiO$_2$ matrix. Energies have been aligned using the states of the embedding SiO$_2$. The impurity atoms are positioned in three different substitutional sites, in the Si-NC center (dot), at the Si-NC/SiO$_2$ interface and in the silica matrix, far away from the Si-NC (silica). The Si atoms at the interface can form three different possible suboxide types, Si$^{1+}$ ,  Si$^{2+}$ , and Si$^{3+}$ depending on the number of bonded O atoms (int-1, int-2, int-3).  For each case, the PDOS of the dopant atom is also reported. Black and grey dots mark the HOMO and the LUMO states, respectively, while the Fermi level  E$_F$ is reported by dashed line. Horizontal lines mark the uncorrected (orange) and corrected (green) band-edge of SiO$_2$.  Reprinted with permission from Ref. \citep{NuriaNS2015}.}
\label{fig:figureexptheo}
\end{center}
\end{figure}

As pointed out before, in recent works \citep{KonigSR2015,HillerSR2017} it has been suggested, based on electronic transport measurements in P-doped Si-NCs, that the impurity atoms are located preferentially in interstitial sites. DFT calculations have been performed for P impurities in SiO$_2$, in SiO$_{0.9}$ and in Si-NC/SiO$_2$ systems, where the SiO$_{0.9}$ and SiO$_2$ were simulated through transition Si-O shells capped by hydrogen. The results showed that P-atoms located in interstitial sites generate states in the fundamental gap of Si-NCs (see Fig. \ref{fig:figureexptheo}). In order to explain the observed quenching of the PL in P-doped Si-NCs, authors hypothesized the occurrence of new non-radiative recombination dynamics generated by these interstitial-P induced states, as an alternative procedure with respect the Auger recombination processes. No calculations were made about the formation energies for the P impurity in this interstitial position. It is worthwhile to note that in a recent study devoted to energetic and carrier transport properties of doped Si-NCs embedded in SiO$_2$ \citep{NuriaNS2015} it was found that P prefers to stay in the Si-NC and B at the interface, bonded to three oxygen atoms and that, in this last case, the formation energy is even lower with respect to the undoped case.

\section{Carrier transport} 
\label{sec:trans}
\begin{figure}
\begin{center}
\includegraphics[width=12.0cm,height=9.43cm]{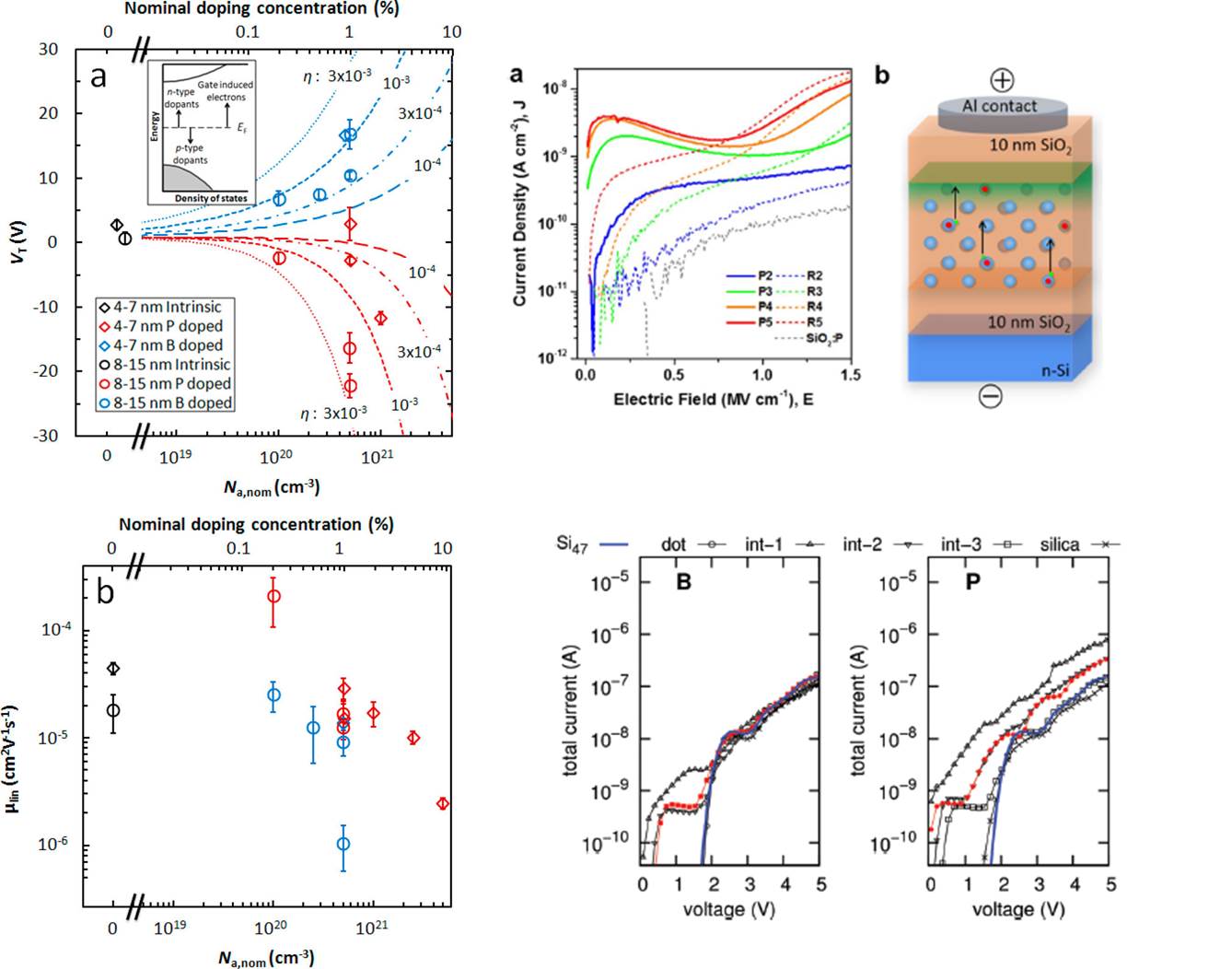}
\caption{(Color online) Left Panel: (a) threshold voltage summary as function of nominal doping concentration. Lines are calculated values for various dopant activation efficiencies. Inset: density of states diagram and the effects of dopants and gate-induced electrons on the Fermi levels. (b) Dopant activation efficiency dependence on nominal doping concentration for (red) P, (black) intrinsic and (blue) B Si-NC thin films for (diamonds) small, 4-7 nm and (circle) large, 8-15 nm, Si-NC thin films.  Reprinted with permission from Ref. \citep{GresbackACS2014}. Top Right Panel: (a) current density vs. electric field for P-doped samples (P: Si-NCs incorporating on average 1 P-atom per Si-NC, R: undoped reference samples). At low E-fields the P-samples (solid lines) clearly exhibit higher displacement current densities than the respective reference samples (dashed lines). (b) The schematic shows the sample structures (large blue spheres: Si-NCs, red points: P-atoms in Si-NCs, small green points: donor electrons localized at P-atoms in thermal equilibrium). Since charge injection from gate or substrate is blocked at low fields, only transient displacement currents caused by P-donors are measured, which accumulate under the gate electrode (shown in green). Reprinted with permission from \citep{HillerSR2017}. Bottom Right Panel: calculated total current (electron + hole) as a function of the applied voltage, for the considered doped configurations (symbols) along with the undoped cases (blue solid curve) for a Si$_{47}$-NC embedded in SiO$_2$. Filled symbols (in red) highlight the most stable configuration. Reprinted with permission from Ref. \citep{NuriaNS2015}.}  
\label{fig:figureexptheo1}
\end{center}
\end{figure}
In this section we will briefly analyze carrier transport in doped Si-NCs, and we will discuss how electric mobility can be used to discern the role played by the impurities. 

In order to prove that  impurities located in Si-NCs effectively act as dopant, it is necessary to assess their electrical activity. This can be done using, for instance, conductivity and resistivity data \citep{ConibeerPPRA2011,LechnerJAP2008,StegnerPRL2008,QianNRL2016}, current density versus electric field measurements \citep{ConibeerPPRA2011,ZhangJAP2015,HillerSR2017,YuSM2015,VeettilAPL2016}, Hall mobility experiments \citep{ShanNano2016} and field effect transistor analysis \citep{GresbackACS2014}. Due to the difficulties in the interpretation of the results obtained in transport measurements, that often depend on a large number of concomitant effects (like for instance transport due to defects, the presence of hyperdoping effects and  transport in percolated nano-Si networks), today little is known about the effective electronic activation. 

From experimental results, it seems possible to conclude that doping depends on synthesis methods and that B and P can dope Si-NCs. However very low dopant activation efficiency were recorded, that is 10$^{-3}$- 10$^{-6}$, where the lower limit refers to B (see Fig \ref{fig:figureexptheo1}, left panel and top right panel). Theoretically, it has been shown that an increment of the conductivity, calculated with respect the  undoped embedded Si-NCs, can be obtained only for P-doping \citep{NuriaPRB2013}. Instead, the electrical response of B-doped embedded Si-NCs does not differ from the undoped case \citep{NuriaNS2015} (see see Fig \ref{fig:figureexptheo1}, bottom right panel).

\section{Codoping}
\label{sec:codo}
The main limitation to high PL efficiency in Si-NCs is due to radiationless Auger recombination. This drawback can be circumvented by simultaneous p-type (B) and n-type (P) compensated doping of the Si-NCs \cite{OssiciniAPL2005,FujiiAPL2004,FujiiJAP2003}.
In this section we will discuss the concept of impurity compensation in Si-NCs. We will show that B and P codoped compensated Si-NCs exhibit properties 
that strongly differ from the ones of the undoped and single doped Si-NCs. Moreover we will prove  that codoping can be exploited to tune the electronic and optical properties of Si-NCs, in a highly controllable way.

Codoped Si-NCs were firstly grown by Fujii and coworkers \cite{FujiiAPL2004,FujiiJAP2003} using a co-sputtering method. During the annealing, the Si-NCs were grown in borophosphosilicate glass (BPSG) matrices. The evidence that impurities were located into substitutional sites of the Si-NCs was obtained through infrared absorption spectra \cite{MimuraPRB2000} and by electron spin resonance (ESR) \cite{FujiiAPL2004,FujioAPL2008}. 

The PL spectra of the B and P codoped Si-NCs strongly differ from the ones recorded for both the single doped (B or P) and the undoped Si-NCs \cite{FujiiAPL2004,MimuraJLum1999,FujiiJAP1998}. The codoped Si-NCs, for instance,  exhibit PL energies red-shifted with respect to those of the corresponding undoped Si-NCs; in particular PL peaks, originated by optical transitions between donor and acceptor states, extend from the visible range to energies below the energy gap of the bulk Si \cite{FujiiAPL2005,FujiiAPL2004,NakamuraPRB2015}. 
\begin{figure}
\begin{center}
\includegraphics[width=5.55cm,height=7.32cm]{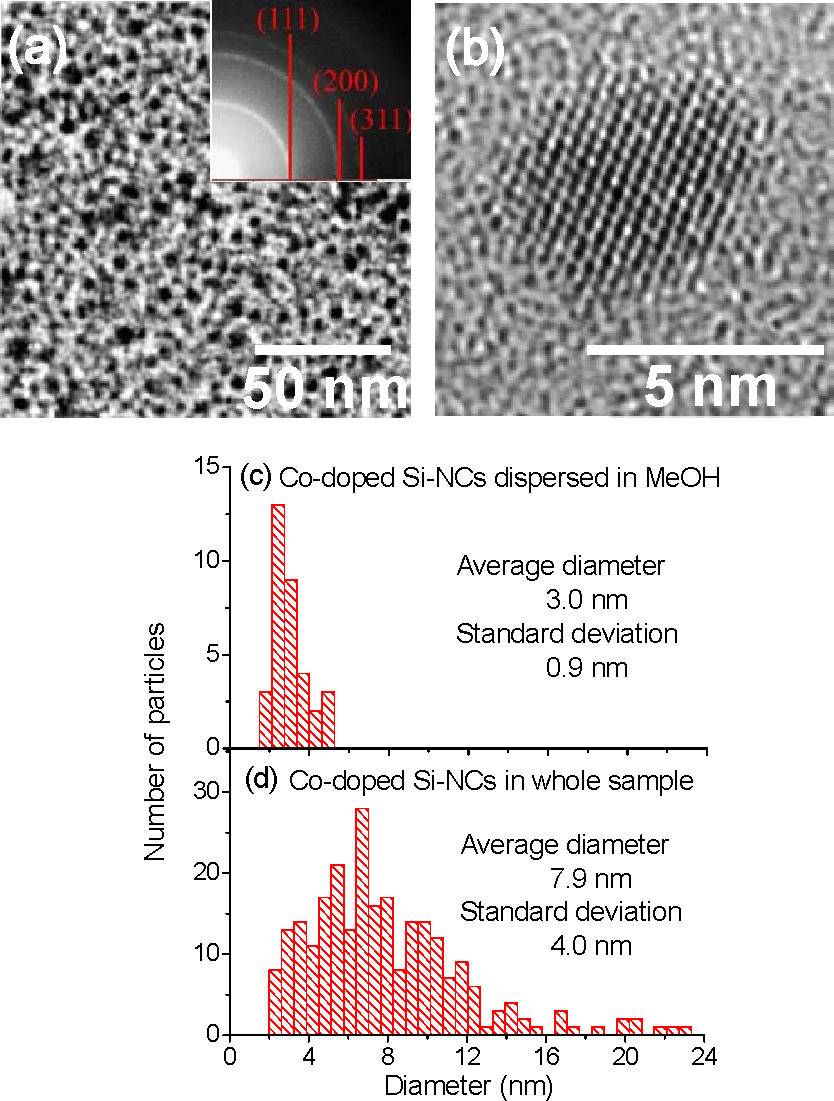}
\caption{(Color online). (a) Bright-field TEM image of P- and B- codoped colloidal Si-NCs. The inset is the electron diffraction pattern, (b) HRTEM image of a P- and B- codoped Si-NC. The lattice fringes correspond to $\{111\}$ planes of Si. The background amorphous image is from a carbon thin film. The size distribution of (c) codoped Si-NCs dispersed in methanol and (d) those in a whole sample containing precipitates. Reprinted with permission from Ref. \citep{SugimotoJPCC2012}.}
\label{figure11b}
\end{center}
\end{figure}

More recently the same group reported a new method to obtain B and P codoped colloidal Si-NCs, where NCs were dispersed in methanol without a surface functionalization process 
\cite{SugimotoJPCC2012,SugimotoJPCC2013a,SugimotoJPCC2013b,KannoJAP2016}. To isolate Si-NCs from BPSG matrices, the BPSG films were dissolved in HF solution in ultrasonic bath. This process produced isolated Si-NCs dispersed in solution, that were separated from the HF by centrifugation obtaining in the concentrator a Si-NC powder. Methanol was then added to disperse the Si-NCs. Even if the majority of Si-NCs were dispersed in methanol, a fraction of Si-NCs were precipitated. These precipitates were removed by centrifugation and a supernatant liquid was obtained \citep{SugimotoJPCC2013a}. 
High resolution transmission electron microscopy (HRTEM) images showed the presence of lattice fringes corresponding to the $\{111\}$ planes of a Si crystal. Moreover it proved the presence  of a large number of defected-free Si-NCs with an average diameter of about 3 nm (see Fig. \ref{figure11b}).
 
\begin{figure}
\begin{center}
\includegraphics[width=7.32cm,height=9.53cm]{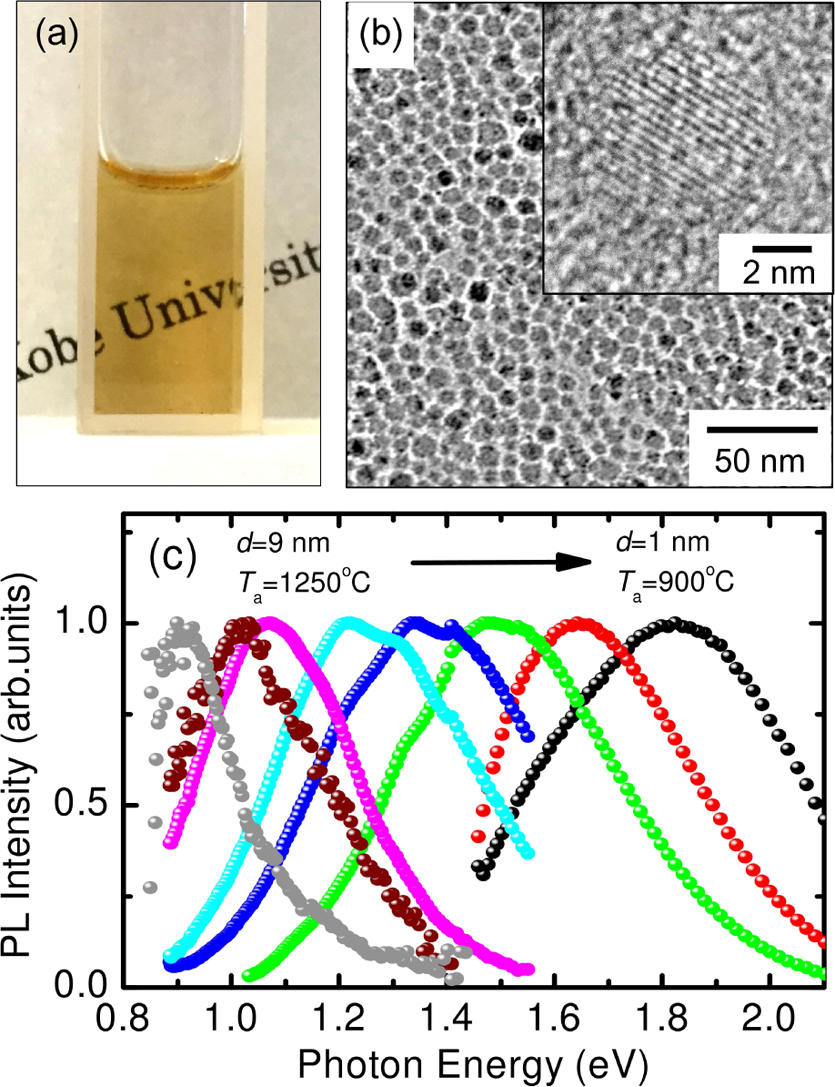}
\caption{(Color online). (a) Photograph of a colloidal dispersion of codoped Si-NCs (methanol solution). (b) TEM image of Si-NCs. Inset is a high resolution images of a Si-NC. (c) PL spectra of codoped Si-NCs with different size, that is, grown at different temperatures. Reprinted with permission from Ref. \citep{HoriNL2016}.}
\label{figure11c}
\end{center}
\end{figure}

An advantage of this method concerns the possibility of producing highly dispersive Si-NCs in polar solvents or water solutions of different PH, without employing organic ligands, thus permitting the use of these Si-NCs for  biomedical applications \citep{OstrovskaRSC2016}. Figure \ref{figure11c} shows a picture of a methanol solution in which codoped Si-NCs of about 5 nm in diameter are dispersed \citep{HoriNL2016}. 

The solution is very clear and light scattering by agglomerates is not seen. The average diameter of the codoped Si-NCs can be varied from 1 to 9 nm controlling the growth temperature and the resulting PL energy peaks range from 1.85 to 0.90 eV. Since after dispersion of the Si-NCs in methanol, the position of the PL peak shifted to slightly higher energy with increasing storage time and then is stabilized, the authors concluded that the surface dangling bonds were terminated by O molecules \citep{SugimotoJPCC2013b}. 

Codoping of Si-NCs has been recently obtained also for silica embedded Si-NCs using ion implantation \citep{NakamuraPRB2015}, co-sputtering \citep{ChungPSSA2016} or using B$_2$H$_6$ and PH$_3$ gases in PECVD  \citep{LiAPL2017}. In all these cases donor-acceptor pair recombination emission has been clearly observed. Moreover evidence of carrier multiplication were detected \citep{ChungPSSA2016}.
  
\subsection{Ab-initio calculations: codoped free-standing and matrix-embedded nanocrystals}
\label{codotheo}
In order to interpret experimental results, we will report the outcomes of ab-initio simulations for  B and P simultaneously codoped Si-NCs 
\cite{OssiciniJNN2008,OssiciniAPL2005,IoriPRB2007,OssiciniJSTQE2006,OssiciniSS2007,
MagriJCMSE2007,GuerraJACS2014,RamosPRB2008,IoriJL2006,IoriPSSA2007,IoriSM2008,IoriPE2009,AmatoJAP2012,AmatoJPD2014}.  

\begin{figure}
\begin{center}
\includegraphics[width=12.cm,height=4.61cm]{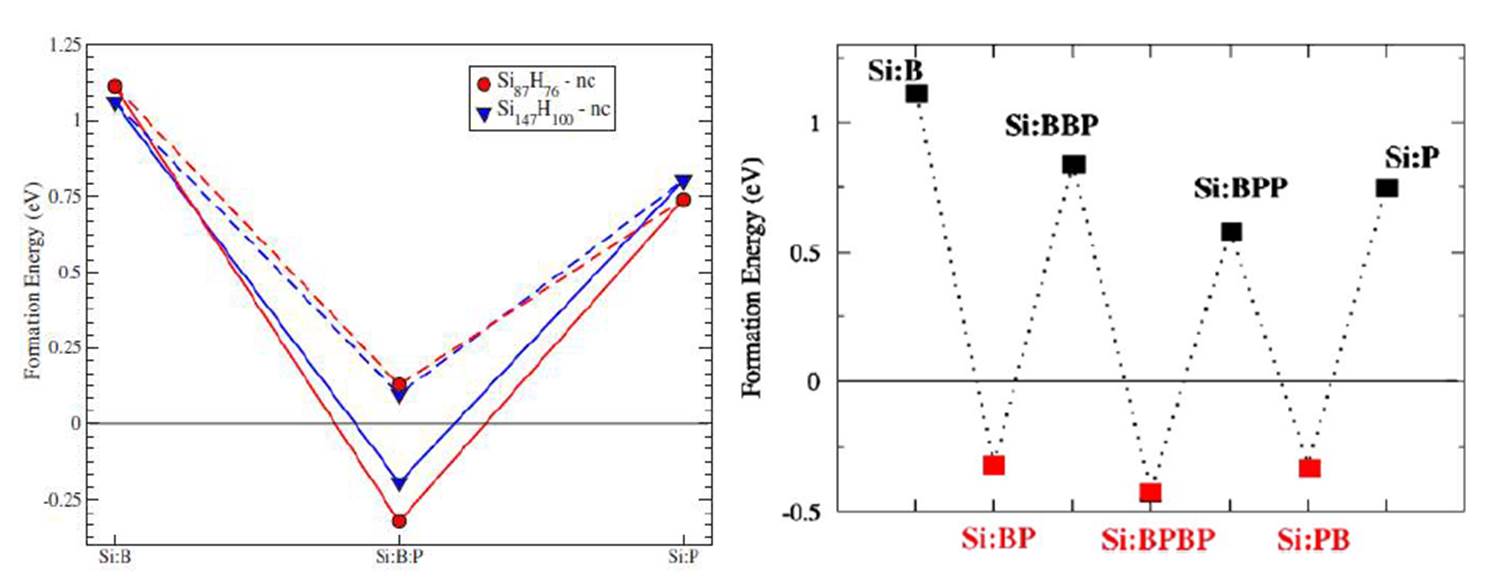}
\caption{%
(Color online) Left Panel: impurities formation energies for single-doped and codoped Si-NCs. Red circles are related to Si$_{87}$H$_{76}$-NC (diameter 1.56 nm) and blue triangles  to Si$_{147}$H$_{100}$-NC (diameter 1.86 nm). Two different impurity-impurity distances are considered in the codoped Si-NCs (dashed and solid lines, larger and smaller distances, respectively).  The lines are guide for the eyes. Reprinted with permission from Ref. \citep{IoriPRB2007}. Righ Panel: formation energis for single, codoped and multi-doped Si$_{147}$H$_{100}$-NC. The lines are guide for the eyes. Reprinted with permission from Ref. \citep{IoriPE2009}.}
\label{figure11}
\end{center}
\end{figure}

Starting from H-terminated Si-NCs, B and P impurities have been located in substitutional positions inside the Si-NCs. Full relaxation of all the obtained structures was then allowed. In Fig. \ref{figure11} (left panel) the calculated impurity formation energies (see Eq. \ref{eq1}) for the Si$_{85}$H$_{76}$-NC and the Si$_{147}$H$_{100}$-NC are reported and compared with the ones calculated for 
B or P single-doped Si-NCs. As an example, in these codoped Si-NCs the dopants are located at two different distances, that is the nearest possible distance between two Si-NC subsurface sites (0.364 nm and 0.368 nm after geometry relaxation for the Si$_{85}$BPH$_{76}$ and the Si$_{145}$BPH$_{100}$ nanocrystals respectively) and the largest possible impurity distance (B and P located in opposite Si-NC subsurface sites, 1.060 nm and 1.329 nm after geometry relaxation for the Si$_{85}$BPH$_{76}$ and the Si$_{145}$BPH$_{100}$ nanocrystals respectively). 
    
From Fig. \ref{figure11} it is clear that simultaneous B and P doping strongly reduces (by about 1 eV) the formation energy with respect to both B or P single doped cases and that this reduction is similar for Si-NCs of different sizes. While B or P single doping is very costly, B and P codoping is much easier, independently on the Si-NC size \cite{OssiciniAPL2005,IoriPRB2007}. Moreover  a reduction of the distance between the two impurities results in a reduction of the formation energy; its minimum is  obtained when the impurities are located at the nearest neighbors positions at the surface of the Si-NCs. In these cases the formation energy assumes negative values. It is interesting to note that Nomoto et al. \cite{NomotoJPCC2016} have observed that codoping is an effective means of promoting segregation and stability of the B and P impurities in the Si-NC region.
The fact that Si-NCs can be more easily codoped than single-doped explains the possibility of easely grow B and P codoped Si-NCs \citep{FujiiAPL2005,FujiiAPL2004,FujiiJAP2003}. In particular, in  the codoped Si-NCs, the differences among the four impurity-Si bond lengths are always smaller with respect to the single-doped case. Thus, if carriers in the Si-NCs are perfectly compensated by simultaneous n- and p-type impurity doping, an almost T$_d$ configuration is recovered; in this configuration the four impurity-Si bonds are practically the same \cite{IoriPRB2007,GuerraJACS2014,RamosPRB2008}. 

These results are confirmed by the calculation performed in the case of multi-doping \citep{IoriPE2009}. In Fig. \ref{figure11} (right panel) the calculated formation energies of Si$_{147}$H$_{100}$-NC are reported as a function of the number of impurities. We note that the presence of an odd number of dopants (1 or 3) brings the formation energies to higher values. Instead, the presence of an even number of compensated B and P impurities (2 or 4) strongly lowers the formation energies, thus indicating that doping is always much easier to realize when one has the same number of donor and acceptor impurity atoms. 

\begin{figure}
\begin{center}
\includegraphics[width=12.cm,height=4.64cm]{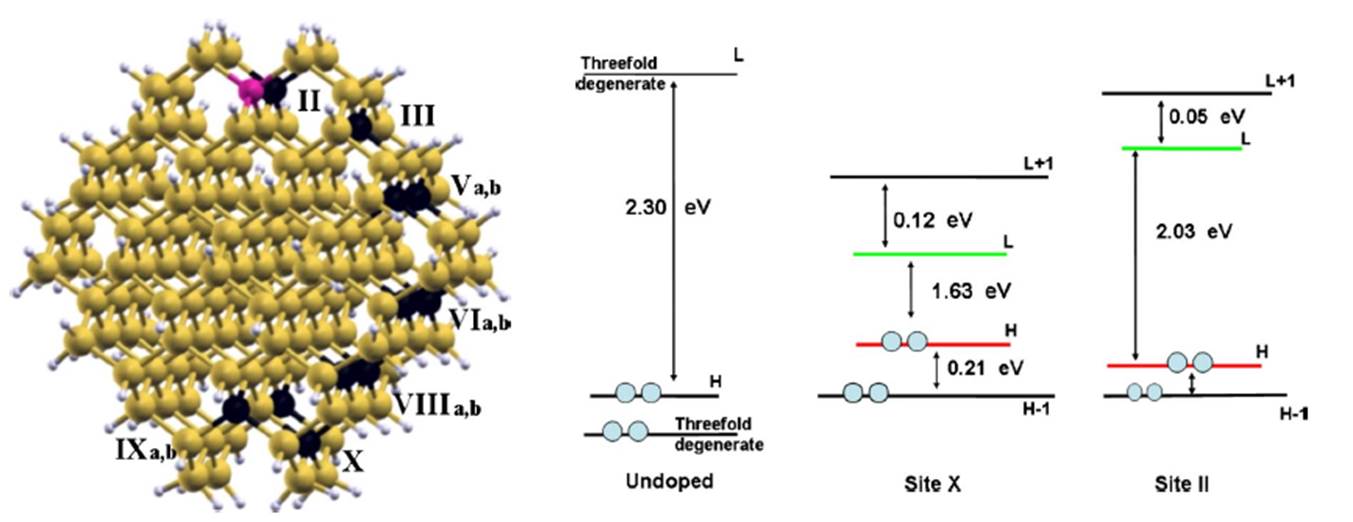}
\caption{%
(Color online). Left: phosphorus impurity path in Si$_{145}$BPH$_{100}$-NC. The B atom (magenta, dark grey) position is kept fixed, while the P atom (black) is moved to explore several substitutional sites, labeled by Roman numbers. The position II corresponds to the minimum distance between the dopants (D = 0.368 nm), whereas the position X is related to the maximum distance (D = 1.329 nm). Right: calculated energy levels of the undoped Si$_{147}$H$_{100}$-NC anf the codoped Si$_{145}$BPH$_{100}$-NC with two different impurity-impurity distances. H stands for HOMO and L for LUMO. Reprinted with permission from ref. \citep{IoriPRB2007}.}
\label{figure12}
\end{center}
\end{figure}

Also the electronic properties of B- and P- codoped Si-NCs are qualitatively and quantitatively different from those of the single doped (B or P) or undoped cases.
The presence of both a n- and a p-impurity leads to the formation of
a HOMO state that contains two electrons and to the formation of a HOMO-LUMO gap that is strongly lowered with respect to the one of the corresponding undoped Si-NCs. The energy levels for the undoped Si$_{87}$H$_{76}$-NC and the codoped Si$_{85}$BPH$_{76}$-NC, calculated at the $\Gamma$ point in the optimized geometries, are shown in the right panel of Fig. \ref{figure12}, where only the levels corresponding to the HOMO, LUMO, HOMO-1 and LUMO+1 states are depicted.
 
\begin{figure}
\begin{center}
\includegraphics[width=12.cm,height=4.5cm]{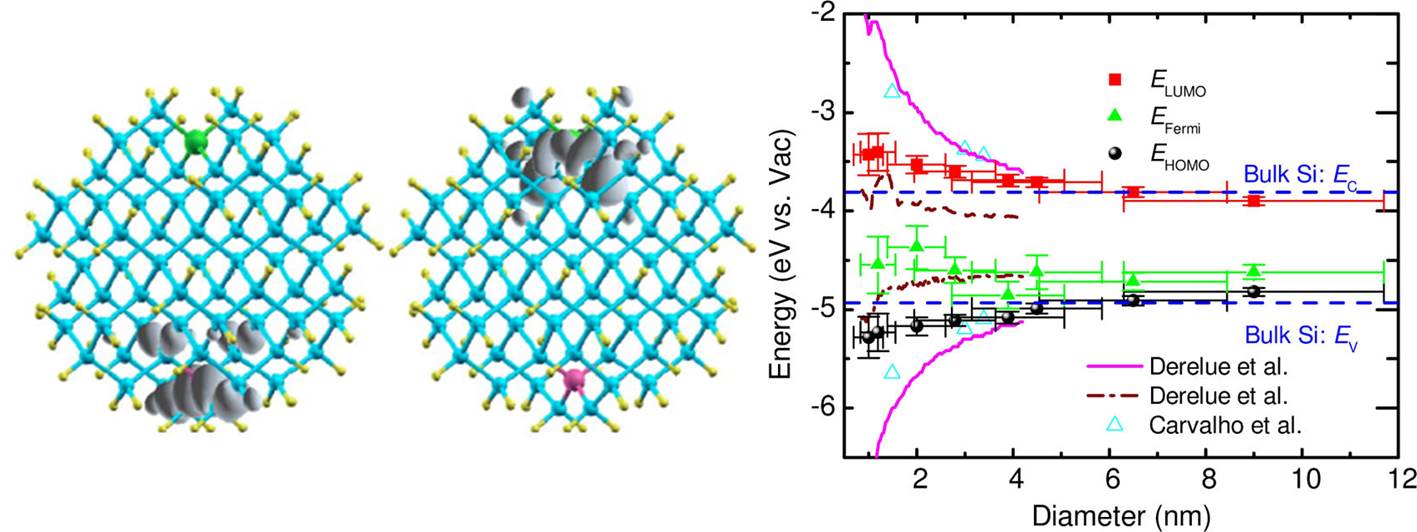}
\caption{%
(Color online) Left: HOMO (left) and LUMO (right) square modulus contour plots calculated for the 
Si$_{85}$BPH$_{76}$-NC. Light blue balls represent Si atoms, while the yellow balls are the H atoms used to saturate the surface danglind bonds. B (violet) and P (green) dopants are here located in subsurface positions on opposite sides of the Si-NC. Reprinted with permission from Ref. \citep{OssiciniAPL2005}. Right: HOMO, LUMO and Fermi level energies measured from the vacuum levels of B and P codoped Si-NCs of different diameters. The horizontal dashed lines are the conduction and valence band edges of bulk Si. Valence band and conduction band edges of undoped Si-NCs calculated by tight binding approximations (solid curves)\citep{DelerueTSF1995} and first principle calculations (open triangle) \citep{CarvalhoPRB2011}. Broken curves are acceptor and donor levels calculated  by tight binding approximations \citep{DelerueTSF1995}. Reprinted with permission from Ref. \citep{HoriNL2016}.}
\label{figure13}
\end{center}
\end{figure}

In Fig. \ref{figure12}, left panel, two different distances between the compensated dopants  are considered, that correspond to the minimum and the maximum separation between impurities located in subsurface positions. From Fig. \ref{figure12} it is clear that in both codoped cases the  HOMO-LUMO energy gap is always lower than the one calculated for the undoped Si-NCs. Besides, when the impurities are at large distances the energy gap is strongly reduced; this reduction is less evident when the impurities are brought closer, as a consequence of the increased Coulomb interaction. These results underlined that, by tuning NCs size and impurity locations, it is possible to obtain codoped Si-NCs with a gap ranging from the visible range to values below the bulk Si energy gap  
\cite{IoriPRB2007,RamosPRB2008,IoriPE2009}, in agreement with the experimental observations \cite{FujiiAPL2005,SugimotoJPCC2013b,KannoJAP2016,FujiiAPL2004,FukudaJL2011}. Results reported in Fig. \ref{figure12} point out that, in codoped Si-NCs, the HOMO and the LUMO states correspond to the acceptor and donor levels of the B and P impurities. This result is confirmed by the plots of Fig.  \ref{figure13},  which clearly demonstrate that HOMO and LUMO wavefunctions  are localized on the B and P impurities, respectively. These results are in good agreement with experimental observations, as show in the right panel of Fig. \ref{figure13} where the HOMO, LUMO and Fermi energies are reported for codoped Si-NCs of different diameters.
The experimental HOMO and LUMO energy levels show a clear size dependence that differs from the one predicted by tight binding for hydrogen-type impurities in H-terminated Si-NCs \citep{DelerueTSF1995}, whereas they are in the same energy range of those calculated by ab-initio methods \citep{OssiciniAPL2005,IoriPRB2007,CarvalhoPRB2011}.  

\begin{figure}
\begin{center}
\includegraphics[width=12.cm,height=7.84cm]{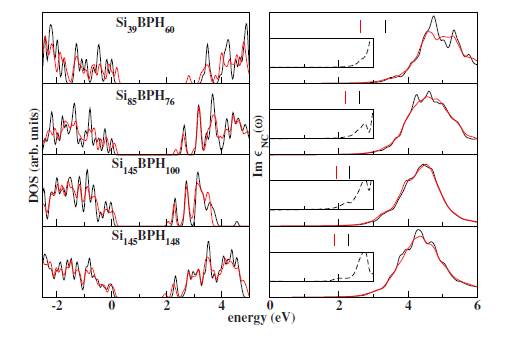}
\caption{%
(Color online) Electronic density of states (left panels) and optical absorption spectra (right panels) for B-P codoped Si-NCs (red lines) in comparison with those of the undoped Si-NCs (black lines). The vertical lines indicate the position of the lowest transition energy. In the right panles, the insets show the difference of absorption intensities between the codoped and the undoped cases (dashed lines) enlarged by a factor 100. Reprinted with permissin from Ref. \citep{RamosPRB2008}.}
\label{fig:figure00}
\end{center}
\end{figure}

The reduction of Si-NC energy gap as a consequence of the simultaneous presence of B and P impurities is also confirmed by  DFT calculations of the density of states (DOS) and of the  absorption spectra (see Fig. \ref{fig:figure00}) \citep{RamosPRB2008}.  By comparing the calculated DOS obtained for the undoped and the codoped cases, we observe the presence of states localized in the forbidden energy region of the undoped NC, states that are originated by the presence of B and P impurities. The energy localization of these new states (that can be experimentally detected by scanning tunneling microscopy or other techniques \citep{WolfNL2013}), and therefore the resulting energy gap, depend on the NC size.

\begin{figure}
\begin{center}
\includegraphics[width=12.cm,height=8.03cm]{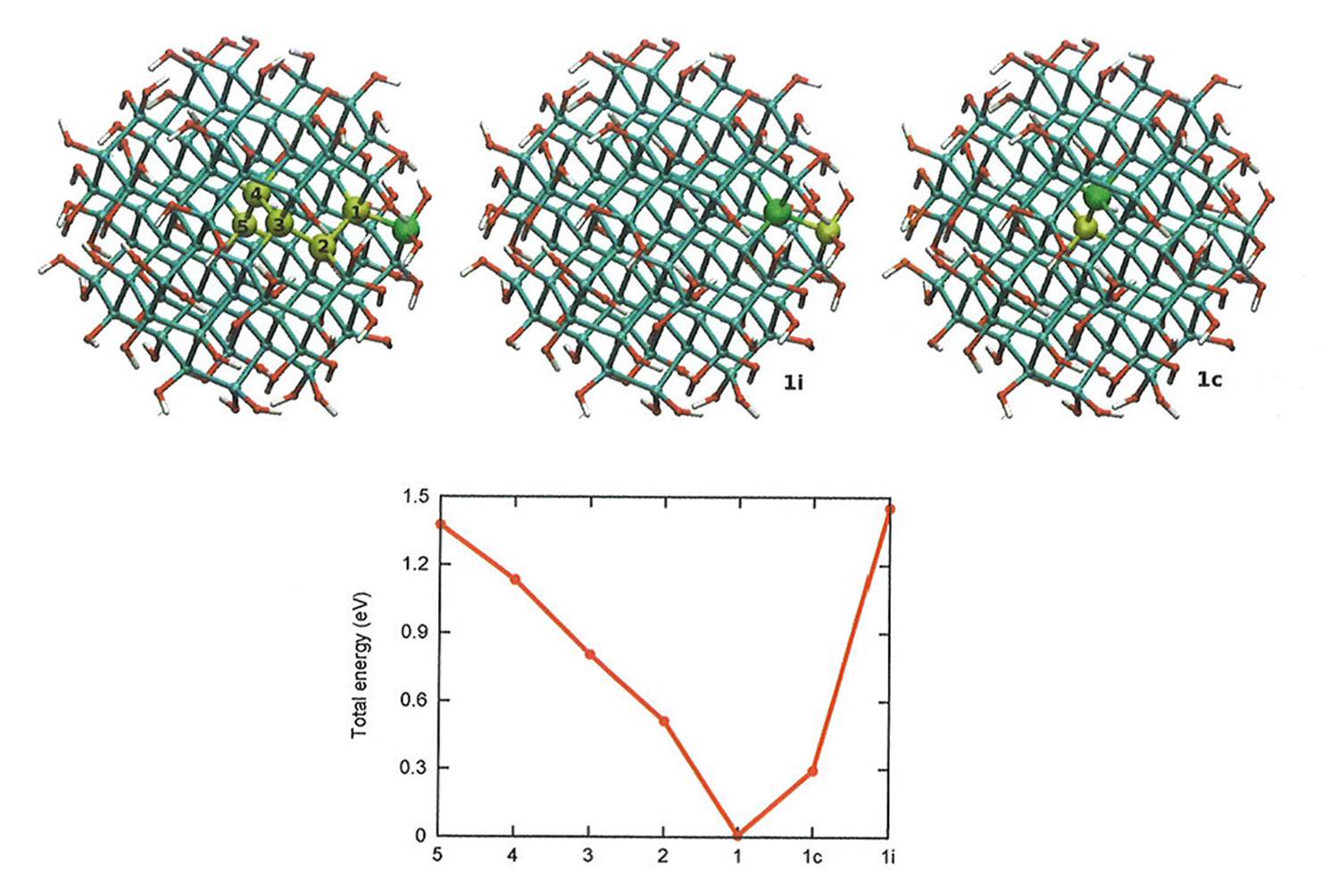}
\caption{%
Top figure: Si$_{85}$BP(OH)${76}$-NC doped (left panel) with B impurity at the surface and P impurity at one of the sites from the center (5) to the nearest neighbor of B (1), doped (central panel) with P placed at the surface and B in the inner nearby (1i); doped (right panel) with P at the Si-NC center and B in the outer nearby (1c). Cyan, red, white, green and yellow balls represent Si, O, H, B, and P atoms, respectively. Bottom figure: total energy of the codoped Si$_{85}$BP(OH)${76}$-NC as function of the dopants position represents in the top figure. Reprinted with permission from Ref. \citep{GuerraJACS2014}.}
\label{figure14}
\end{center}
\end{figure}

Indeed, a free-standing codoped Si-NC, the Si$_{85}$BP(OH)$_{76}$, was generated and studied by Guerra et al. \citep{GuerraJACS2014}. As a first step, impurities were located by considering their preferred position in the single doped Si-NCs, that is by placing P in the centre of the Si-NC and B at the Si-NC surface. Then  structural and electronic properties were investigated as a function of the impurities separation by progressively approaching P to B. In Fig. \ref{figure14} the considered systems are schematised. Here the P positions are enumerated from 5 (Si-NC center) to 1 (Si-NC inner surface).  The calculated total energies are reported in the botton part of Fig. \ref{figure14}. It is evident that while in the single-doped Si-NC the P atom stabilizes at the Si-NC center, in the codoped case the formation of a strong P-B bond (position 1) is clearly energetically favored.

Alternatively, the P-B bond can be formed at the Si-NC center, fulfilling the "minimum enegry costrain rule" of the single-doped Si-NC for the P dopant and forcing the B atom in the Si-NC core. Such configuration, named 1c, results more energetically favored than the 5, 4, 3, and 2 configurations, confirming
the idea that the formation of the P-B bond prevails with respect to other parameters.
However, the configuration of position 1 (P-B pair at the interface)  remains most stable than the configuration of position 1c (P-B pair at the centre).

Due to the polar nature of the P-B bond, the above condition implies the formation of a static electric dipole radially directed and preferentially located at the Si-NC surface, which points toward the Si-NC center. Similar results have been obtained by Ni et al. \citep{NiPRB2017}. The high stability of the dipole direction is evidenced by considering a configuration in which the codopants are switched, a B-P pair, named 1i (see Fig. \ref{figure14}). Quite interestingly, such configuration produces a total energy higher than all the other cases. The formation of stable dipoles at the Si-NCs surfaces radially directed inward the Si-NCs can provide an explanation of the production of stable colloidal codoped Si-NCs without surface functionalization processes
\citep{SugimotoJPCC2012,SugimotoJPCC2013a,SugimotoJPCC2013b,SugimotoNS2014}. Moreover, since in the codoped case the minimum energy formation is found for the strong and stable P-B bond, also the intensity of the transition between HOMO (acceptor) and LUMO (donor) states in the codoped Si-NCs is affected by the impurity distance. Stronger transitions arise when the impurities are closer, whereas the intensity gets lower when the impurities are at larger distances, as shown by the calculated absorption spectra \citep{IoriPRB2007}. These results were confirmed by the experimental determination of strong PL intensities in colloidal and matrix-embedded codoped Si-NCs \citep{SugimotoJPCC2013b}. Similar theoretical outcomes have been obtained also for codoped Si-NWs \citep{AmatoJAP2012,AmatoJPD2014,AmatoCR2014,AmatoPSS2016}. 

\begin{figure}
\begin{center}
\includegraphics[width=12.cm,height=6.72cm]{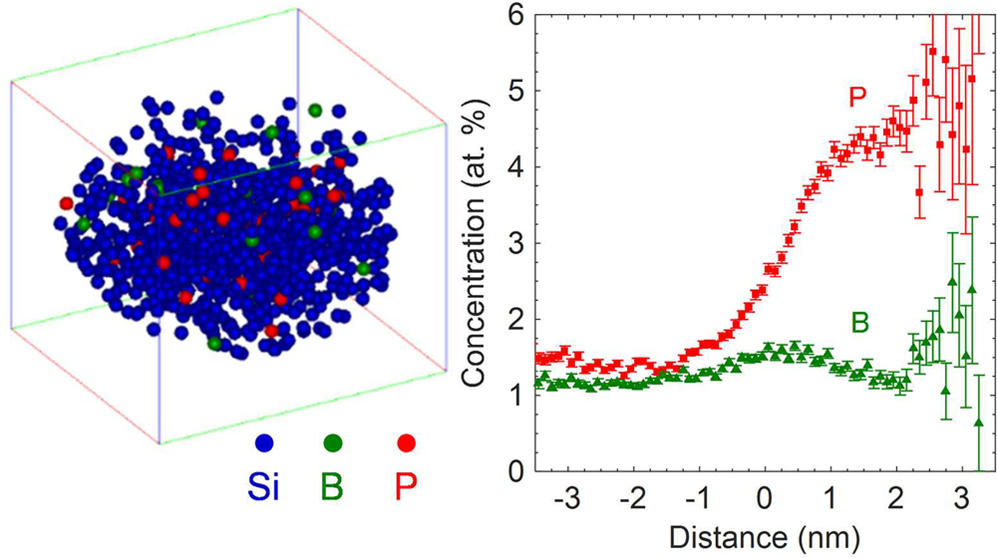}
\caption{%
(Color online). Individual Si-NC in BPSG annealed at 1150 °C. The diameter is 4.0 nm. On the left the relative  proxigram analysis. The distance at 0 nm represents the interface between the Si-NC and the matrix. The positive distances represent the inside of the Si-NC, while the negative distances represent the matrix.  Reprinted with permission from Ref. \citep{NomotoJPCC2016}.}
\label{figure11a}
\end{center}
\end{figure}

All these results found a confirmation in the atom probe tomography analysis of the B and the P distribution in codoped Si-NCs \citep{NomotoJPCC2016}. In Fig. \ref{figure11a} the proxigram analysis of codoped Si-NCs in BPSG matrix annealed at 1150 $\degree$C is reported. The distance at 0 nm represents the interface between the Si-NC and the matrix. Going to positive (negative) distance means moving towards the center of the Si-NC (inside the matrix). Clearly in the codoped Si-NCs the P concentration increases on going into the Si-NC, i.e. the P atom prefers to stay inside the Si region, whereas the B concentration does not show abrupt changes and reaches its maximum exactly at the Si-NC/matrix interface. Moreover a cluster analysis found that a large fraction of the B and P atoms present in the codoped samples were consumed to form B-P clusters in the Si-NCs, in good agreement with the theoretical results shown in Fig. \ref{figure14}.

\section{Conclusions}
\label{sec:concl}
We presented a review of the recent activities concerning doping in silicon nanocrystals. Particular attention has been devoted to the theoretical outcomes that often complement the experimental analysis.
After two decades of intense efforts, there is consensus about the possibility of incorporating
 boron and phosphorus as impurities in Si-NCs as small as 2 nm in diameter, where  quantum confinement effects are particularly important. It is also clear that the incorporation of B and P in the Si-NCs does not depend only on the Si-NC size, but it is also influenced by the Si-NC interface and by the embedding medium. Theoretical and experimental outcomes pointed out that there are important differences between free-standing and matrix-embedded Si-NCs. 
A rapidly growing effort is devoted to the characterization of compensated codoped Si-NCs. Here we expect interesting results in the next years and an important impact in the development of future optoelectronic devices. For what concern doping in Si-NCs the
 most complicated issue regards the dopant electronic activation. Even if it has been demonstrated that this activation is possible, the efficiency is very low for both p- and n- impurities, with a clear preference for phosphorus with respect to boron. There is then space for new ideas aiming at the development of innovative strategies to realize doping in Si-NCs.

\section{Acknowledgements} 
We are grateful to M. Amato, F. Iori and R. Guerra for useful discussions. The authors acknowledge the CINECA award under the ISCRA initiative, for the availability of high performance computing resources and support and the PRACE for awarding us access to resource Marconi and FERMI based in Italy at CINECA.
Ivan Marri acknowledges support/funding from European Union H2020-EINFRA-2015-1 programme under grant agreement No. 676598 project "MaX - materials at the exascale".

\section{References}
\bibliography{sidoping}
\end{document}